\documentstyle[seceq,mbf,wrapft]{ptptex}

\pubinfo{Supplement no.~133, 1999}  
\preprintnumber[3cm]{
KUNS-1325\\PTPTeX ver.0.8\\ August, 1997}

\title{
Gravitational Lensing in Clusters of Galaxies
}

\author{
Makoto {\sc Hattori},$^{1,}$\footnote{E-mail address: hattori@astr.tohoku.ac.jp} 
Jean-Paul {\sc Kneib}$^{2,}$\footnote{E-mail address: jean-paul.kneib@ast.obs-mip.fr} and
Nobuyoshi {\sc Makino}$^{3,}$\footnote{E-mail address: makino@astro.oita-ct.ac.jp}  
}

\inst{
$^1$ Astronomical Institute, Tohoku University, Sendai 980-8578, Japan
\\
$^2$ OMP, 14 Av. E. Belin, 31400 Toulouse, France
\\
$^3$ 
Oita National College of Technology,
1666 Maki, Oita 870-0152, Japan
}

\recdate{March 12, 1999      
}

\abst{
Gravitational lensing in clusters of galaxies is an efficient tool to probe the
mass distribution of galaxies and clusters, high
redshift objects thanks to the gravitational amplification, and the
geometry of the universe. We  review some important
aspects of cluster lensing and related issues in {\it observational}
cosmology.
}

\begin{document}

\maketitle


\makeatletter
\if 0\@prtstyle
\def\asp{.3em} \def\bsp{.26em}
\else
\def\asp{.3em} \def\bsp{.3em}
\fi \makeatother

\section{Introduction: Discovery of  giant luminous arcs}

\begin{wrapfigure}{r}{6.6cm}
\vspace{6.3cm}  
\special{epsfile=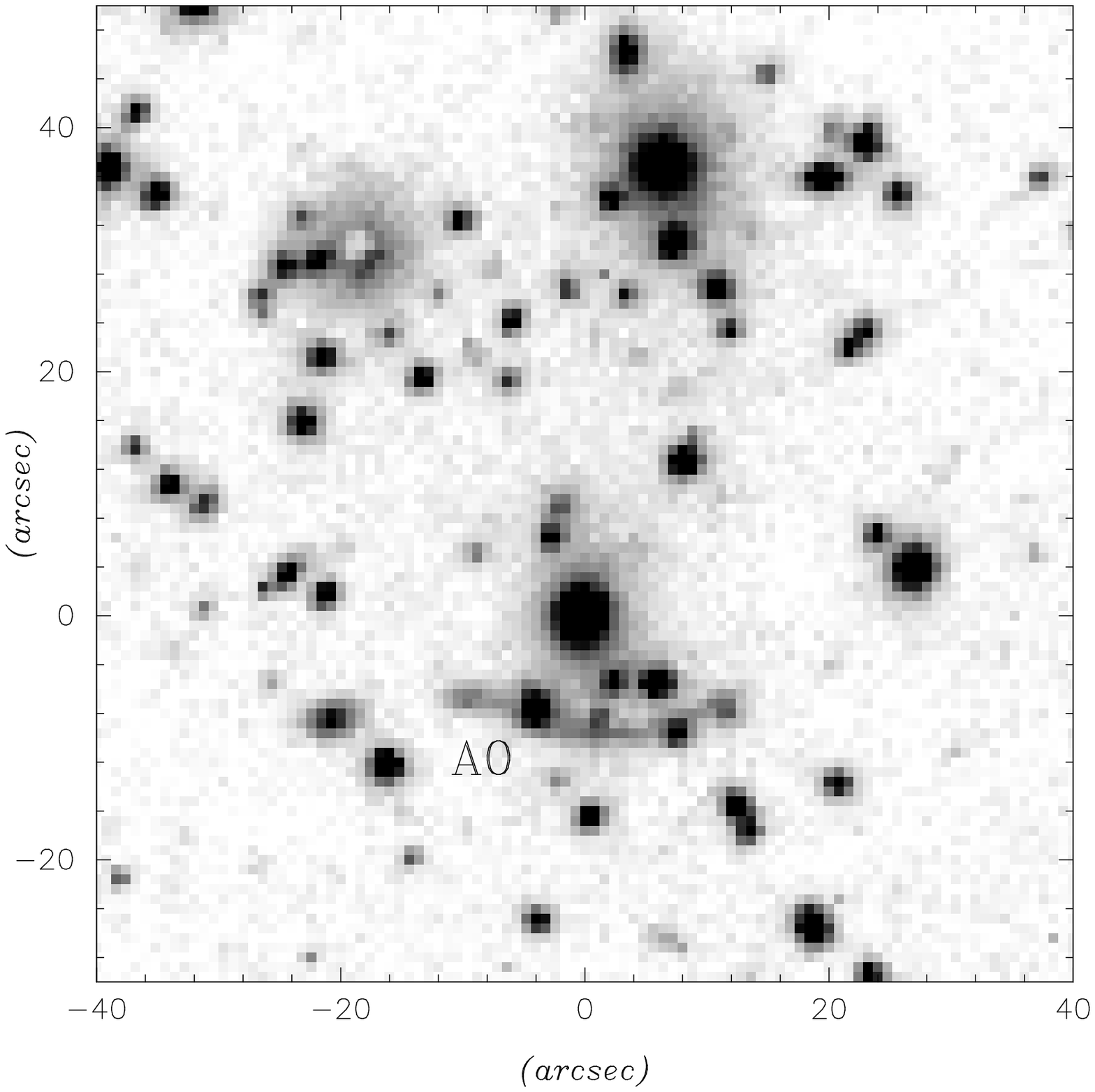  hoffset=-10 voffset=200 hscale=0.4 vscale=0.4}
\caption{The discovery of the Giant Luminous Arc.  An arc-like structure (A0)
appeared on the CCD image of central part of 
cluster A370 at $z=0.374.$\cite{Soucail87} The image was taken during autumn, 1985, with a 
$320\times512$ pixels CCD using the Canada-France-Hawaii-Telescope (CFHT). 
The pixel size was $\sim 0.8$ arcsec. The strange  blotch at 
position $\sim(-20,30)$ is a CCD defect in the original
CCD.}
\label{fig:a370old}
\end{wrapfigure}

More than a dozen years ago, the discovery of extremely elongated and
luminous arc-like images in three distant rich clusters of galaxies
A370, A2218 and CL2244-02 was reported by Lynds and
Petrosian\cite{Lynds86} in the Bulletin of the American Astronomical
Society.  Independently, Soucail et al.\cite{Soucail87} (see
Fig. \ref{fig:a370old}) discovered many arcs in the cluster A370
($z=0.37$).  Soon after, Soucail et al.\cite{Soucail88} measured a
redshift of $z=0.724$ for the giant arc in A370 --- nearly twice the
redshift of the lensing cluster --- thus confirming the gravitational
nature of the arc, as suggested by Paczy\'nski.\cite{Paczynski87}
Giant luminous arcs (GLA) are typically bluer, longer and generally
brighter than (normal) cluster galaxies.

These discoveries have promoted gravitational lensing to 
the position of a major
cosmological tool for the following purposes:  i) study the cluster mass distribution;
ii) probe high redshift galaxies using clusters as natural
gravitational telescopes;  iii) constrain the geometry of the
universe, as parameterized by $\Omega_0$ and $\Lambda_0$, 
and especially the value of the cosmological constant; and
iv) search for large-scale structures in the
high redshift universe.  This emerging new area seems to represent 
the realization of 
Zwicky's\cite{Zwicky37a,Zwicky37b} dreams. 

In this paper, we present the status of  gravitational lensing 
studies of clusters of galaxies in regard to the above 
four topics ---  except the status of the weak lensing mass
reconstruction, which is reviewed in another paper in this volume.\cite{ume99}

This review is organized as follows. Section 2 gives the fundamentals
of gravitational lensing in order to understand how the cluster mass
distribution can be constrained in the strong lensing regime. This
description is restricted to key equations, since several very good
reviews on lensing fundamentals have been published.  The reader
who is not familiar with the field of gravitational lensing can
consult the reviews by Blandford and Kochanek,\cite{Blandford87}
Blandford and Narayan,\cite{Blandford92} Schneider, Ehlers and 
Falco,\cite{SEF92} Fort and Mellier,\cite{Fort94} Wu\cite{Wu96} and
Narayan and Bertelmann\cite{NB97} and Mellier.\cite{Mellier98}  Section
3 presents an overview of the modelling of the mass distribution in
clusters in the strong lensing regime, with a 
comparison between the lensing and X-ray mass estimates.  Indeed, a
discrepancy of a factor of 2--3 is generally observed between strong
lensing and de-projection of the X-ray emission of the intra-cluster
medium (henceforth ICM).  Proposed solutions to this problem are
summarized.  Section 4 discusses the ``dark'' lens problem in
connection to the search for very high redshift clusters.  
Section 5 presents recent applications
of the use of clusters of galaxies as natural gravitational telescopes
to study the nature of high redshift galaxies.  Section 6 summarizes
the studies connecting the statistics of lensed galaxies in clusters, nicknamed
``arc statistics".  Arc statistics are sensitive to:  i) the
cosmological world models (and particularly to $\Omega_0$ and
$\Lambda_0$); ii) the average cluster mass distribution/profile;
and  iii) the nature and evolution of high redshift galaxies.
Clearly, arc statistics depend on a large number of parameters. We
therefore attempt here to clarify its possible application in order to
sharpen focus of considerations on 
its future use.  Section 7 summarizes the proposed ideas to
constrain the geometry of the universe by using the lensing effect due
to clusters of galaxies.  In each section, we attempt to explain how the
new generation of telescopes and instruments (SUBARU, VLT, Gemini,
Astro-E, AXAF, XMM, NGST, LSA/MMA, LMSA ...) may enrich the
cluster-lensing subject.

This review focuses on various aspects (particularly those we
have been deeply involved with) and should not be considered 
exhaustive.  However, we believe that this focus makes this review unique from
others and that it will help readers to catch up quickly with the
current frontier of this subject.

\section{Fundamentals of cluster lensing}

Gravitational lensing theory is based on the limit of a weak
stationary gravitational field.  Furthermore, in the case of cluster
lensing, the size of the deflector is much smaller than the light
propagation distance.  Therefore, the lens can be safely approximated
by a single plane of matter with a projected surface mass density
$\Sigma (\vec{\theta})$ --- usually referred as the {\it thin lens
  approximation}. All our discussion is based on these
assumptions.

\subsection{Definition of an {\it arc}}

General relativity states that the presence of massive object
distorts  space-time. Thus  propagating light beams coming
from distant sources will be sensible to any mass inhomogeneities
`en route'. The light coming from any extended sources and passing
through a massive cluster of galaxies will therefore suffer major
distortions and amplifications due to the gravitational lensing effect.
In some extreme cases, the distortions are non-linear (on the scale of the
image), and the image is strongly curved, having
no doubt as to the gravitational nature of the image.
Such lensed images are usually called  ``arc", yet it is
clearly a subjective definition.  In this paper, to quantify the
definition of a lensed image, we use the term {\it arc} for images
with an axis ratio (length-to-width) larger than 10, and {\it arclet} for
images with length-to-width ratios between 2 and 10.  Furthermore,
giant arcs brighter than $V=22.5$ are defined as {\it giant luminous
arc} (GLA), following the definition of Le F\'evre et al.\cite{Lefevre94}
 
\subsection{Lensing equation: Mapping and deformation}

The lensing equation can simply be expressed as a mapping from the 
image plane to the source plane.
For an angular position of the image $\vec{\theta}$,
the angular position of the source $\vec{\beta}$ 
is given by the lensing equation
\begin{equation}
\vec{\beta}=\vec{\theta}-\vec{\nabla_{\theta}}\psi(\theta),\label{eq:1}
\end{equation}
where $\psi(\theta)$ is the effective lensing potential
and $\vec{\nabla_{\theta}}$ is the 2D gradient by $\vec{\theta}$.
It is important to realize that this equation may admit multiple solutions
in $\vec{\theta}$ for a fixed $\vec{\beta}$, depending on the values of
the potential $\psi$. If multiple solution exists, we will say that the
lens produces multiple images of a source at position $\vec{\beta}$; this
domain defines the strong lensing regime (see below).
 
The effective lensing potential $\psi$
is defined by the Poisson equation,
\begin{equation}
\vec{\nabla_{\theta}}^2\psi =2{\Sigma(\vec{\theta})\over \Sigma_{\rm 
cr}}\equiv 2\kappa(\vec{\theta}),\label{eq:2}
\end{equation}
where $\Sigma_{\rm cr}$ is the critical surface mass density defined by 
\begin{equation}
\Sigma_{\rm cr}={c^2\over 4\pi G}{D_{\rm s}\over D_{\rm d}D_{\rm
ds}}.\label{eq:3}
\end{equation}
The surface mass density normalized by the critical  surface mass density
is called the `convergence' $\kappa(\vec{\theta})$.
The effective lensing potential can be obtained by solving the Poisson 
equation (\ref{eq:2}):
\begin{equation}
\psi(\vec{\theta})={1\over \pi}\int\kappa(\vec{\theta'}){\rm
ln}|\vec{\theta} -\vec{\theta'}|d^2\theta'.\label{eq:4}
\end{equation}

A surface element of the image plane $\delta\vec\theta$ is related
to a surface element of the source plane $\delta\vec\beta$ via the 
inverse of the magnification matrix $A$, defined by
\begin{equation}
\delta\vec\beta= 
{\partial \vec\beta \over\partial \vec\theta} \delta\vec\theta
=\left({I -\vec\nabla_\theta\vec\nabla_\theta \psi} \right)\delta\vec\theta
=A^{-1} \delta\vec\theta,
\label{eq:jp1}
\end{equation}
where $\vec\nabla_{\theta}\vec\nabla_{\theta}$ is $2\times2$ diadictensor
and $I$ is $2\times 2$ unit matrix.
The magnification matrix can
be written as a function of the second derivatives of the potential.
In Cartesian coordinates, it reads
\begin{equation}
A^{-1} = \left(
             \begin{array}{cc}
              1 - {\partial^2\psi \over \partial\theta_x^2} & 
                      -{\partial^2\psi \over \partial \theta_x\partial\theta_y} \\
              {\partial^2\psi \over \partial \theta_x\partial \theta_y} &
               1 - {\partial^2\psi \over \partial \theta_y^2} \\
              \end{array}
         \right),
\end{equation}
and in polar coordinates,
\begin{equation}
A^{-1} =  \left(
              \begin{array}{cc}
              1 - {\partial^2\psi \over \partial \theta^2} & 
                      -{\partial \over \partial \theta}
                       \left({1\over\theta}{\partial\psi \over \partial \varphi}\right) \\
               -{\partial \over \partial \theta}
                       \left({1\over\theta}{\partial\psi \over \partial \varphi}\right)&
               1 -{1\over \theta} {\partial\psi \over \partial \theta}
                 -{1\over \theta^2}{\partial^2\psi \over \partial \varphi^2} \\
              \end{array}
          \right),
\end{equation}
where $\theta=\vert\vec{\theta}\vert$, and $\varphi$ is the azimuthal angle of the image
position.  Note surface elements of image and source planes should be taken 
as $\delta\vec{\theta}=(\delta\theta,\theta\delta\varphi)$ 
and $\delta\vec{\beta}=(\delta\beta,\beta\delta\varphi')$
in the polar coordinates, where $\beta=\vert\vec{\beta}\vert$ and 
$\varphi'$ is the azimuthal angle of the source position. 

In the limit of a single lens plane, the magnification matrix
is real and symmetric. Thus in this limit it is diagonalizable.
The transformation of a surface element is non-linear and can
be expressed by two effects, an isotropic magnification expressed
by  the convergence $\kappa$ and a deformation expressed by
the complex shear $\overline\gamma$:
\begin{equation}
A^{-1} = \left( 
              \begin{array}{cc}
              1 - \kappa - \gamma_1 & \gamma_2 \\
                          \gamma_2  & 1 - \kappa + \gamma_1 \\
              \end{array}
         \right),
\label{eq:jp2}
\end{equation}
where $\overline\gamma = \gamma_1 + i \gamma_2
                       = \gamma (\cos(2\theta_A) + i \sin(2\theta_A))$
with $\theta_A$ giving the shear direction. 

In the shear coordinates, the magnification matrix is diagonal
with proper values $1-\kappa-\gamma$ and $1-\kappa+\gamma$.
The total magnification $\mu$ is defined by the determinant of the
magnification matrix (when it is defined):
$\mu = \det A = 1/((1-\kappa)^2-\gamma^2)\equiv H(\vec{\theta})^{-1}$,
where $H(\vec{\theta})$ is the Hessian of the lens mapping.
Depending on the sign of the total magnification  $\mu$, 
the parity of the image will change. 
For $\mu>0$, the image has the same parity as the original source, even, while
for $\mu<0$, the image has  
mirror-symmetry compared to the original source, and the image parity
is {\it odd}.
The family of points for which the inverse of the magnification matrix
is singular (infinite magnification: $\mu^{-1}=0$)
are called {\it critical lines}. They correspond to
$\kappa+\gamma=1$ (external) or $\kappa-\gamma=1$ (internal).
The corresponding curves in the source plane are called
{\it caustic lines}.

For a non-uniform mass distribution, if
$\Sigma(\theta)$ exceeds $\Sigma_{\rm cr}$ in a region of the image
plane, it ensures that the equation $\mu^{-1}=0$ admits a solution. Thus 
critical lines exist.  Clusters with a surface mass density
higher than the critical surface mass density are called {\it
critical} lenses, and such a surface mass density is said to be {\it
super critical}.
 
Arcs occur close to and across critical lines 
(as the merging of two or more ``multiple images'').
They are strongly deformed  and highly magnified. 
The region in which  arcs can be found is referred to 
as the  {\it strong lensing} regime.

\subsubsection{Case of a circular symmetric lens}

Although the circular symmetric case is an unrealistic description of a
cluster mass distribution, it is helpful to understand the
gravitational lensing. We shall discuss it in more detail here.

Indeed, in the case of a circular symmetric lens, the lens equation reads
\begin{equation}
\beta({\theta})=\theta - {M(\theta)\over \pi D_{\rm d}^2\theta\Sigma_{\rm
cr}},\label{eq:5}
\end{equation}
and the inverse of the magnification matrix is (in polar coordinates)
\begin{equation}
A^{-1} = \left(\begin{array}{cc}
               {d\beta\over d\theta} & 0 \\
               0  & {\beta\over\theta} 
               \end{array}
         \right)
       = \left(\begin{array}{cc}
               1-{1\over \pi D^2_{\rm d}\Sigma_{\rm cr}}
                {\partial\over\partial\theta}
                \left({M(\theta)\over\theta}\right) & 0 \\
               0  & 1-{1\over \pi D^2_{\rm d}\Sigma_{\rm cr}}
                \left({M(\theta)\over\theta^2}\right)
               \end{array}
         \right),
\label{eq:jp3}
\end{equation}
where $M(\theta)$ is the lens mass enclosed within a radius $D_{\rm
d}\theta$,
\begin{equation}
M(\theta)=2\pi D_{\rm
d}^2\int_0^{\theta}\Sigma(\theta')\theta'd\theta'.\label{eq:6}
\end{equation} 

The factor $\theta\over\beta$ in (\ref{eq:jp3}) is the tangential deformation factor of
the image.  This can easily be understood by geometrical
considerations.  In the case of a circular symmetric lens, a light ray
can only be radially displaced.  Now, consider a (small) source at
position $\beta$ from the centre of the mass distribution with a
length $l$ in the tangential direction. Its image is located at
position $\theta$ with a tangential length $l' =
l\times({\theta\over\beta})$, as only radial displacements are allowed.
Similarly, $({d\beta\over d\theta})^{-1}$ is the deformation factor of
the image in the radial direction, because this is the ratio of the
unlensed source width, $\delta\beta$, to the lensed image width,
$\delta\theta$.

When the surface mass density follows a power law of the form
$\Sigma(\theta)=C_0\theta^{-\delta}$ with $\delta<2$, 
the magnification matrix can be written as 
\begin{equation}
A^{-1} = \left(\begin{array}{cc}
 1-{2C_0\over \Sigma_{\rm cr}}\left({1-\delta\over 2-\delta}\right)
   {1\over \theta^\delta} & 0 \\
 0  &  1-{2C_0\over \Sigma_{\rm cr}}{1\over 2-\delta}
   {1\over \theta^\delta}
\end{array}
\right).
\label{eq:jp4}
\end{equation}

In the circular case, critical lines are easily defined,  by 
symmetry considerations; they are circles. The tangential (external) critical
curve given by $\beta/\theta=0$ is a circle, called an
`Einstein ring'. The Einstein radius is defined by
\begin{equation}
\theta_{\rm E}={1\over D_{\rm d}}
\left({M(\theta_{\rm E})\over \pi\Sigma_{\rm cr}}\right)^{1/2}.
\label{eq:7}
\end{equation}
The radial deformation factor at the Einstein radius is
\begin{equation}
\left({d\beta\over d\theta}\right)^{-1}=
               \left[2\left(1-{\Sigma(\theta_{\rm E})\over
                     \Sigma_{\rm cr}}\right)\right]^{-1}.
\label{eq:rad}
\end{equation}
When the surface mass density follows a power law of the form
$\Sigma(\theta)\propto\theta^{-\delta}$, 
the radial deformation factor can be written as 
\begin{equation}
\left({d\beta\over d\theta}\right)^{-1}={1\over\delta}.\label{eq:radel}
\end{equation}
Therefore, for a lens with $\delta=1$, which corresponds to the SIS 
model, the radial deformation factor is unity, while a more compact mass
distribution ($\delta>1$) forms narrower arcs (the radial stretching rate 
is less than 1), and a shallower mass distribution ($\delta<1$) produces 
thicker arcs (the radial stretching rate
is larger than 1). 

The radial (internal) critical curve given by 
$d\beta/d\theta =0$ is either
a point (if the mass distribution is singular) or a circle
of radius 
\begin{equation}
\theta_{\rm R}={1\over D_{\rm d}}
 \left({M(\theta_{\rm R})\over\pi \Sigma_{\rm cr}}\right)^{1\over 2}
\left({1\over{2\Sigma(\theta_{\rm R})\over\Sigma_{\rm cr}}-1}\right)^{1\over2}.
\label{eq:8}
\end{equation}
When $\Sigma(\theta)\propto\theta^{-\delta}$, the convergence and the
tangential deformation factor at the radial critical radius are given
by
\begin{equation}
\kappa(\theta_{\rm R})={1\over 2}
\left({2-\delta\over 1-\delta}\right),\label{eq:radconv}
\end{equation}
\begin{equation}
\left\vert{\theta\over \beta}\right\vert=
\left\vert 1-{1\over\delta}\right\vert.\label{eq:radfac}
\end{equation}
Equation (\ref{eq:radconv}) shows that $\delta<1$ is required for a lens to
have a radial critical line, since the convergence must be positive.
Equation (\ref{eq:radfac}) shows that for a lens with $\delta=0.5$, the
tangential deformation factor is unity, while a more compact mass
distribution ($1>\delta>0.5$) forms thinner arcs and a shallower mass
distribution. The case $\delta<0.5$ produces thicker arcs.

If the source position is slightly off of centre, the ring is
split into two tangential highly stretched arc lying close to the
Einstein ring.  The typical angular separation of images is of order
$2\theta_{\rm E}$. Therefore, if a pair of arcs or multiply imaged
arcs are found in the cluster core, the Einstein radius can be
estimated from the distance between the arc and cluster centre, and a
very crude estimation of the mass of the cluster within the Einstein
radius can be estimated by
\begin{equation}
M(\theta_{\rm E})=\pi\theta_{\rm E}^2\Sigma_{\rm
cr}.\label{eq:8}
\end{equation}
The critical surface density $\Sigma_{\rm cr}$ is obtained by
measuring the  redshifts of the lens cluster, $z_{\rm d}$, and source,
$z_{\rm s}$, by fixing the values of $\Omega_0$, $\Lambda_0$ and $H_0$. 
For example, for $z_{\rm s}=1$ and $z_{\rm d}=0.3$ with
($\Omega_0$,$\Lambda_0$,$H_0$)$=$(1.0,0.0,$100h{\rm km/sec/Mpc}$),
we have
\begin{equation}
\Sigma_{\rm cr}=1.09h^{-1}{\rm g/cm^2},\label{eq:9}
\end{equation}
which is of the order of the weight of a sheet of paper.

Before the discovery of giant arcs, the Coma cluster was 
the archetype of a rich cluster of galaxies.  
The central surface mass density of the Coma estimated 
from its X-ray emission, assuming spherical symmetry
and isothermal hydrostatic equilibrium, is given by 
\begin{equation}
\Sigma_{\rm Coma} (0)= 
{3\over 2G} \beta {k_{\rm B}T\over \mu m_{\rm H}}=0.37\left({kT\over 
8.3{\rm keV}}\right)\left({\beta \over 0.75}\right)\left({r_c\over
200h^{-1}{\rm kpc}}\right)^{-1}{\rm g\;cm^{-2}},\label{eq:10}
\end{equation}
where the electron density is assumed to be described by 
$n_e(r)=n_e(0)(1+({r\over r_c})^2)^{-{3\over 2}\beta}$.
Since this value is smaller than the critical surface mass density,
it was believed that clusters of galaxies could not produce
strong lensing events. Clearly, the discovery of giant arcs 
has revolutionized the view we had of the mass distribution
of cluster cores.

\subsubsection{Spherical lens models}

The analytical formulae of frequently used spherically symmetric mass
distribution models are summarized in this subsection.\\

\noindent$\bullet${\it Singular isothermal sphere (SIS) model}

A singular isothermal sphere (SIS) is a solution of the
collisionless Boltzmann equation. The word ``isothermal'' means here
that the velocity dispersion of the system is isotropic and uniform.  SIS is
frequently used in lensing analysis, since the density profile of SIS
is very simple, and most quantities related to gravitational
lensing are described in simple analytic forms. It is instructive to
examine lensing properties using SIS (see, for example, Turner,
Ostriker and Gott III\cite{TOG84}).

The mass density of SIS is described by
\begin{eqnarray}
  \rho(r)&=&\rho_0\left(\frac{r_0}{r}\right)^2\nonumber\\
         &=& \frac{\sigma^2}{2\pi G r^2},
\end{eqnarray}
where $r_0$ is the scaling parameter, $\rho_0$ is the mass density 
at $r_0$, and $\sigma^2$ is the line of sight velocity dispersion.

Integrating the above equation along the line of sight, one obtains the
surface density of the singular isothermal sphere at an angle
$\theta$,
\begin{equation}
  \Sigma(\theta)=\frac{\sigma^2}{2GD_{\rm d}\theta}.
\end{equation}
The lens mass enclosed within a radius $D_{\rm d}\theta$, is found to be
\begin{equation}
  M(\theta)=\frac{\pi\sigma^2 D_{\rm d}\theta}{G}.
\end{equation}
Although the mass within a finite radius does not diverge, the
central density and the total mass diverge. Therefore this model
is not a physical representative of a realistic mass distribution
and should only be considered as a simple model to quickly
estimate physical parameters.\\ 

\noindent$\bullet${\it Isothermal sphere with a finite core radius}

To avoid the divergence found in SIS, a finite core can be added.
The density profile of an isothermal sphere with a finite core radius
is therefore defined by
\begin{equation}
  \rho(r)=\rho_0\left(1+\frac{r^2}{r_c^2}\right)^{-1},
\end{equation}
where $\rho_0$ is the density at the centre, and $r_c$ is the core
radius.

The surface density of an isothermal sphere with a finite core radius is
\begin{equation}
\Sigma(\theta)=\frac{\pi\rho_0D_{\rm d}\theta_c^2}{\sqrt{\theta^2+\theta^2_c}},
\end{equation}
where $\theta_c=r_c/D_{\rm d}$.
The mass enclosed within the radius $\theta$ is
\begin{equation}
  M(\theta)=2\pi^2D_{\rm d}^3\rho_0\theta_c^2
  \left(\sqrt{\theta^2+\theta_c^2}-\theta_c\right).
\end{equation}

Similarly as the SIS, lensing quantities of an isothermal sphere with a
finite core radius are also described by simple analytic
functions. However, the total mass of this model is still infinite.\\

\noindent$\bullet${\it Modified Hubble law}

The modified Hubble law is used as an analytical approximation of the
King model, which is one of the solutions of the collisionless Boltzmann
equation. The King model can be approximated at small radius
by the modified Hubble law.

The mass density profile of the modified Hubble law is given by
\begin{equation}
  \rho(r)=\rho_0\left(1+\frac{r^2}{r_c^2}\right)^{-3/2},
\end{equation}
where $\rho_0$ is the density at the centre, and $r_c$ is the core radius. 
Integrating the density distribution along the line of sight,
one can obtain the surface density
\begin{equation}
  \Sigma(\theta)=\frac{2\rho_0D_{\rm d}\theta^3_c}{\theta^2+\theta_c^2}.
\end{equation}
The projected mass of the modified Hubble law is
\begin{equation}
  M(\theta)=2\pi D_{\rm d}^3\theta_c^3\rho_0\log
  \left(\frac{\theta^2+\theta_c^2}{\theta_c^2}\right).
\end{equation}

Since the mass density vanishes in the limit $r\rightarrow\infty$
faster than SIS, it is difficult to
produce strong lensing events for a finite mass unless it is extremely compact
and massive. The modified Hubble model was
conventionally used in describing a galaxy distribution in a cluster.\\

\noindent$\bullet${\it Isothermal $\beta$-model}

It is empirically known that (generally) the radial profile of the 
X-ray surface brightness can be  described well by the 
so-called $\beta$-model among X-ray astronomers, that is,
\begin{equation}
S_x(r)=S_0\left[1+{r^2\over r_c^2}\right]^{-3\beta+1/2}.
\end{equation}
If the intra-cluster gas is isothermal (the gas temperature is constant
throughout the cluster), and in hydrostatic equilibrium,
the electron density profile is obtained from this formula as 
$n_e(r)=n_e(0)(1+(r/r_c)^2)^{-3\beta/2}$. 
Then we can solve the hydrostatic equation to
obtain the cluster mass distribution. The mass density
distribution of the isothermal $\beta$-model is given by
\begin{equation}
  \rho(r)=\rho_0\frac{3+(r/r_c)^2}{[1+(r/r_c)^2]^2},
\end{equation}
where $r_c$ is the core radius.  The parameter $\rho_0$ is determined
from X-ray observations by fitting the $\beta$-model to the X-ray
surface brightness. It is given by
\begin{equation}
  \rho_0=\frac{3\beta k_{\rm B}T}{4\pi G\mu m_{\rm H}r_c^2},
\end{equation}
where $\beta$ is the slope parameter in the $\beta$-model, 
$k_{\rm B}$ is the Boltzmann constant, $T$ is 
the temperature of the X-ray emitting gas, $\mu$ is the mean molecular weight,
and $m_{\rm H}$ is the Hydrogen mass.
The surface density is
\begin{equation}
  \Sigma(\theta)=\pi\rho_0D_{\rm d}\frac{\theta_c^2(2\theta_c^2+\theta^2)}
{(\theta^2_c+\theta^2)^{3/2}}.
\end{equation}
The projected mass of the isothermal $\beta$-model is 
\begin{equation}
  M(\theta)=2\pi^2D_{\rm d}^3\rho_0\theta_c^2\frac{\theta^2}
    {\sqrt{\theta^2_c+\theta^2}}.
\end{equation}

The temperature required to have an Einstein ring radius of $\theta_{\rm E}$ 
is written as
\begin{equation}
k_{\rm B}T={1\over 6\pi\beta}\mu m_{\rm H}c^2
{D_{\rm s}\over D_{\rm ds}}\sqrt{\theta_{\rm E}^2+
\theta_c^2}.\label{eq:kT}
\end{equation}
This formula is useful to check whether X-ray results 
are consistent with strong lensing mass estimates, 
assuming that the mass distribution is spherical.\\

\noindent$\bullet${\it NFW profile}

The NFW universal density profile is a fit of a {\it relaxed system}
simulated by {\it N}-body simulations with a high
resolution.\cite{NFW} According to Navarro, Frenk and White,\cite{NFW}
all profiles found in the simulations have the same shape,
independent, of the halo mass, initial density fluctuation spectrum,
and values of the cosmological parameters. (For these reasons it was
called `universal'.)

The NFW density profile is given by
\begin{equation}
  \rho(r)=\frac{\rho_{\rm cr}\delta_c}{\frac{r}{r_s}\left(1+\frac{r}{r_s}\right)^2},
\end{equation}
where $\rho_{\rm cr}$ is the critical density, $\delta_c$ is the
density parameter, and $r_s$ is the scaling parameter.

One obtains the surface density by integrating the mass density along the
line of sight. The surface mass density is, for
$\theta<\theta_s(=r_s/D_{\rm d})$,
\begin{equation}
  \Sigma(\theta)=2\rho_0D_{\rm d}\theta_s
 \left[
   \frac{2\theta^3_s}{(\theta^2_s-\theta^2)^{3/2}}
     {\rm arctanh}\sqrt{\frac{\theta_s-\theta}{\theta_s+\theta}}-
   \frac{\theta^2_s}{\theta^2_s-\theta^2}
 \right],
\end{equation}
for $\theta=\theta_s$,
\begin{equation}
  \Sigma(\theta)=\frac{2}{3}\rho_0\theta D_{\rm d},
\end{equation}
and for $\theta>\theta_s$,
\begin{equation}
  \Sigma(\theta)=2\rho_0D_{\rm d}\theta_s\left[
  -\frac{2\theta^3_s}{(\theta^2-\theta^2_s)^{3/2}}
  \arctan\sqrt{\frac{\theta-\theta_s}{\theta+\theta_s}}
  +\frac{\theta^2_s}{\theta^2-\theta^2_s}
 \right],
\end{equation}
where $\rho_0=\delta_c\rho_{\rm cr}$.  For $\theta/\theta_s<10^{-3}$,
$\Sigma\propto\theta^{-0.1}$.  For $10^{-3}<\theta/\theta_s<10^{-1}$,
$\Sigma\propto\theta^{-0.4}$. For $10^{-1}<\theta/\theta_s< 0.3$, 
$\Sigma\propto\theta^{-0.6}$.
For $0.3<\theta/\theta_s< 1$, 
$\Sigma\propto\theta^{-0.95}$.
For $1<\theta/\theta_s< 10$, 
$\Sigma\propto\theta^{-1.5}$.
For $10<\theta/\theta_s$, 
$\Sigma\propto\theta^{-2}$.

The projected mass for $\theta<\theta_s$ is 
\begin{equation}
  M(\theta)=4\pi\rho_0D_{\rm d}^3\theta^3_s\left[\log\frac{\theta}{2\theta_s}+
  \frac{2\theta_s}{\sqrt{\theta_s^2-\theta^2}}{\rm arctanh}
  \sqrt{\frac{\theta_s-\theta}{\theta_s+\theta}} \right],
\end{equation}
for $\theta=\theta_s$,
\begin{equation}
  M(\theta)=4\pi\rho_0D^3_{\rm d}\theta^3_s\left(\log\frac{\theta}{2\theta_s}
  +1\right),
\end{equation}
and for $\theta>\theta_s$,
\begin{equation}
  M(\theta)=4\pi\rho D_{\rm d}^3\theta_s^3\left[
  \log\frac{\theta}{2\theta_s}+\frac{2\theta_s}{\sqrt{\theta^2-\theta^2_s}}
  \arctan\sqrt{\frac{\theta-\theta_s}{\theta+\theta_s}} \right].
\end{equation}
The formulae for the NFW density profile are found in Bartelmann\cite{BART96}
and Maoz et al.\cite{MAOZ97}

\subsubsection{Pseudo-elliptical lensing potential}

If one would like to construct a physically motivated cluster mass
model, one should start from the 3D mass distribution. However,
as long as the mass distribution of the lens is thin enough to be 
described sufficiently well by the {\it thin lens approximation}, only the surface mass
density affects the lensing phenomena.  Furthermore, it is usually
complicated to construct the projected mass distribution and its lensing
properties' model based on a realistic 3D mass distribution even for an
ellipsoidal mass distribution (see the next subsection).  Therefore, the
elliptical surface mass density distribution and the elliptical
lensing potential have been widely used in lens modelling
because they provide simple analytic descriptions of physically
motivated mass models.  In this subsection, the nature of the
non-singular pseudo-elliptical lensing potential Eq. (\ref{eq:pse}) is
summarized, since it has been widely used for lens modelling, and
the behavior of the arcs produced by the pseudo-elliptical lensing
potential illustrates typical elliptical lensing configurations.

The non-singular pseudo-elliptical lensing potential is defined by
\begin{equation}
\psi(\vec\theta)={D_{\rm ds}\over D_{\rm s}}4\pi
{\sigma_v^2\over
c^2}[\theta_0^2+(1-\epsilon)\theta_1^2+(1+\epsilon)\theta_2^2]^{1/2},
\label{eq:pse}
\end{equation}
where $\epsilon=(a^2-b^2)/(a^2+b^2)$ is the ellipticity of the
potential. The ellipticity of the projected mass density is $\sim
3\epsilon$ for small $\epsilon$.\cite{MFK93}  When $\epsilon=0$ and
$\theta_0=0$, $\sigma_v$ is exactly the line-of-sight velocity
dispersion (1D velocity dispersion) of a singular isothermal sphere,
and it is a constant throughout the cluster.

However, in the case of a finite core radius, interpretation of
$\sigma_v$ requires caution.  In the circular potential limit
($\epsilon=0$), $\sigma_v$ can be related to the 1D central velocity
dispersion of the original 3D mass distribution,\cite{MFK93}
$\sigma_0$, by assuming an isotropic velocity distribution,
self-gravitating spherical equilibrium system such as
\begin{equation}
\sigma_0^2={2\over 3}\sigma_v^2. \label{eq:14}
\end{equation}
These are related to the central line-of-sight velocity
dispersion,\cite{MFK93} $\sigma_{\rm los}$:
\begin{equation}
\sigma_{\rm los}^2(0)={9\over 8}\sigma_0^2={3\over
4}\sigma_v^2.\label{eq:15}
\end{equation} 
They are also related to  X-ray observables\cite{HAT98} by
\begin{equation}
\sigma_0^2= {2\over 3}\sigma_v^2 =\beta {k_{\rm B}T\over \mu m_{\rm
H}},\label{eq:sig0}
\end{equation}
where the isothermal $\beta$ model is used for modelling the cluster mass
distribution.

The non-singular pseudo-elliptical lensing potential should be used
with caution, because for 
$\vert\theta_2\vert\gg\vert\theta_1\vert\gg\theta_0$ 
the mass-isodensity contours are
convex only if $\vert\epsilon\vert<1/5$. For $\epsilon>1/5$, the
mass-isodensity has a complex shape, something like a dumbbell.
\cite{Kovner89}

The ellipticity of the mass distribution of 
a self-gravitating non-rotating collisionless system 
(which is likely to be the case for cluster of galaxies)
is limited by the dynamical instability,\cite{MH91}
and the axis ratio must be smaller than $a/b \sim 2.5$. 
Thus we expect that in many cases the elliptical potential cannot
be a good representation of a physically motivated mass distribution.

\subsubsection{Elliptical mass distribution}

In this subsection, 
we first consider  the lensing due to an oblate spheroidal as an example of
a 3D ellipsoidal mass distribution of the lens.\cite{Bourassa73}
The density distribution of the oblate spheroidal is described 
by $\rho(a)$, where $a$ is defined by
\begin{equation}
a^2\equiv x'^2+y'^2+{z'^2\over 1- e^2} \;\;\;{\rm for\;\;1>{\it e}\ge 0}.
\label{eq:18}
\end{equation}
When the symmetry axis $z'$ titled relative to the line of sight by an angle
$\gamma$, the 2D surface mass density is described by
\begin{equation}
\Sigma(x_i,y_i)=\sqrt{1-e^2\over 1-e^2{\rm sin^2}\gamma}
\int_{b^2}^{r_{\rm cut}^2}
da^2{\rho(a)\over\sqrt{a^2-b^2}},
\label{eq:19}
\end{equation}
where $r_{\rm cut}$ is the cut off radius of the cluster mass distribution
in the direction of the major axis of the oblate spheroid,
$x_i$ and $y_i$ are coordinates in the lens plane, which is perpendicular 
to the line of sight, and $b$ is defined by
\begin{equation}
b^2\equiv x_i^2+{y_i^2\over 1-e^2{\rm sin^2}\gamma}.\label{eq:20}
\end{equation}
The complex formulation of the lensing equation is convenient in this case, and 
the lensing equation is written as
\begin{equation}
Z_s=Z-{1\over \pi}I^*(Z),\label{eq:21}
\end{equation}
where $Z_s\equiv x_s+iy_s$ is the source position, $Z\equiv x_i+iy_i$
is the image position in complex form, and $I^*$ is the complex
conjugate of the scattering function, $I(Z)$, which is defined by
\begin{equation}
I\equiv\int dx'_i dy'_i\kappa(b'){1\over Z-Z'}\\
=2\pi{\sqrt{Z^2}\over Z}\sqrt{1-e^2{\rm sin^2}\gamma}
\int_0^{{\rm min}(r_{\rm cut},b)}
db'{b'\kappa(b')\over\sqrt{Z^2-b'^2e^2{\rm sin^2}\gamma}}.\label{eq:22}
\end{equation} 
A more detailed derivation of this equation is given by Asano and
Fukuyama.\cite{asano98}  This result shows that only the mass within
the isodensity contour at the position of the image, $b$, contributes
on the lensing equation.\cite{asano98}

For example, an image positioned on the minor axis will be sensitive
to the mass within an ellipse and thus with a larger area compared to
the circular symmetric case.  Therefore, an elliptical lens need not 
be super critical to have multiple images (see the next section).

We now discuss the properties of the pseudo-isothermal elliptical
mass distribution, since it is one of the most useful elliptical
surface density profiles, and has been extensively applied to the
simple mass estimate of lens cluster.\cite{KK93}  The convergence
of the pseudo-isothermal elliptical mass profile is given by
\begin{equation}
  \kappa(b)=\frac{\kappa_0 b_0}{\sqrt{b^2_0+b^2}},
\label{eq:notruncate}
\end{equation}
where $\kappa_0$ is the convergence at the centre, and $b_0$ is the
core radius.  One obtains the scattering function $I$ from
Eq. (\ref{eq:22}) for Eq. (\ref{eq:notruncate}) with $r_{\rm
  cut}\rightarrow\infty$:
\begin{equation}
 I(Z)=2\pi\kappa_0\theta_c\frac{\sqrt{Z^2}}{iZ}\sqrt{\frac{1-e^2\sin^2\gamma}
{e^2\sin^2\gamma}}\ln\left[\frac{i\sqrt{(b^2_0+\theta^2_c)
e^2\sin^2\gamma}+\sqrt{Z^2-b^2_0e^2\sin^2\gamma}}{
i\theta_c\sqrt{e^2\sin^2\gamma-\sqrt{Z^2}}}\right].
\end{equation}
From the scattering function we can then easily compute the lensing
properties of this mass distribution. Although this mass distribution
does not have a finite total mass, a truncated version of that model
can be constructed (see Kneib et al. \cite{Kneib96}).  The lens
properties of the pseudo-isothermal elliptical mass profile are similar
to those of the pseudo-elliptical lensing potential, in particular the
behavior of image configurations.\cite{KK93}  However, there are
several qualitative differences between these two.  A
pseudo-isothermal elliptical mass distribution can have naked cusps
for any value of the central convergence or the core radius, whereas
a physically meaningful pseudo elliptical lens potential can only have
naked cusps when $\kappa_0<5$. Since the convergence is given in the
pseudo-isothermal elliptical mass profile, the contour of the
isodensity is convex for all values of the ellipticity, and the surface
density is always positive (which is not always the case for an
elliptical potential).

\subsection{Multiple images and time delay: Folds, cusps, lips and beak-to-beak caustics}

When a light ray propagates through a deep gravitational field, the
arrival time of the light to an observer is delayed by two effects:
i) a geometrical effect and ii) a gravitational effect.
The geometrical delay is explained by the deflection of the light by
the gravitational field. The gravitational delay is due to the fact
that a deep gravitational potential is equivalent to a medium with a
refractive index $(1+2\phi/c^2)^{-1}$, which is larger than 1 ($\phi$
is the 3D gravitational potential defined so that at infinity
$\phi$=0). Hence, the effective speed of light propagating through a
deep gravitational potential is ``slower'' and results in a delay of the
arrival time.

For a source at position $\vec{\beta}$ in the source plane, the time
delay at position $\vec{\theta}$ in the image plane is given
by\cite{Blandford86}
\begin{equation}
t(\vec{\theta})={(1+z_{\rm d})\over c}{D_{\rm d}D_{\rm s}\over D_{\rm
ds}} \left[{1\over
2}(\vec{\theta}-\vec{\beta})^2-\psi(\vec{\theta})\right],
\label{eq:11}
\end{equation}
where $z_{\rm d}$ is the redshift of the lens, $D_{\rm s}$, 
$D_{\rm d}$ and $D_{\rm ds}$ are angular diameter distances of the
observer to the source, the observer to the lens, and the lens to the source, 
respectively, and
$\psi(\vec{\theta})$ is the effective gravitational lensing potential,
which corresponds to the gravitational time delay. The effective
gravitational lensing potential is defined by Eq. (\ref{eq:4}), and its
amplitude depends both on the mass distribution of the lensing object
and the distance to the source.  The term 
${1\over  2}(\vec{\theta}-\vec{\beta})^2$ corresponds to the geometrical
effect.  Fermat's Principle states that a light ray follows a
trajectory such that the light-travel time is stationary relative to
neighboring trajectories.  In other words, the images are located at
those points where the time-delay surface $t(\vec{\theta})$ is
stationary: $\vec{\nabla}_{\theta} t(\vec{\theta})=0$. This defines
the lensing equation (\ref{eq:1}), where $\vec{\nabla}_{\theta}$ is
the 2D gradient by $\vec{\theta}$.

The number of stationary points in the time-delay surface corresponds
to the number of expected images.  This number depends on the source
position in the sky, the lensing parameter, which depends on the
redshift of the lens and the source and depth and shape of the lens
potential, and ($\Omega_0,\;\Lambda_0$) through angular diameter
distances.  The behavior of the time-delay surface is described well
by the catastrophe theory.\cite{PS78} In the case of lensing theory,
the number of independent control parameters is three, e.g., two
parameters for specifying the source position in the sky and one lensing
parameter.  Therefore,  elementary catastrophes that play a role
in lensing are folds, cusps, swallow tails, hyperbolic umbilics and
elliptic umbilics.

\begin{figure}
\vspace{5.5cm}  
\special{epsfile=fig2a.ps rotation=-90 hoffset=130 voffset=157 hscale=0.5 vscale=0.5}
\special{epsfile=fig2b.ps  rotation=-90 hoffset=400 voffset=170 hscale=0.5 vscale=0.5}
\caption{The caustics  and critical lines of a circular lens with 
a finite core.  Left and right panels describe the caustics in the source plane
and the critical lines in the image plane, respectively. 
A tangential (radial) caustic is a central point  
in the source plane, 
and the corresponding
critical line is a outer circle in the image plane.  
The appearance of the images
corresponding to each source position are also shown.  }
\label{fig:2}
\end{figure}

For a circular symmetric lens, the nature of the time-delay potential
depends only on the radial distance of the source position from the
lens centre.  In other words, the number of the control parameters is
one.  Therefore, Tom's catastrophe theorem implies that only the fold
catastrophe appears in a circular symmetric lens (Fig. 2).  For
an elliptical lens, the time-delay potential depends on the 
two-dimensional positions of the source, and hence the number of control
parameters is two. Therefore, the cusp and fold catastrophe can appear
in this case (Fig. 3).  When a source crosses a fold caustic from
inside to the outside, two images with opposite parity will become closer,
then merges at the location of the critical line and finally disappear,
when the source moves outside of the caustics.  When a source crosses
a cusp caustic from the inside to the outside, three images with
opposite parity will become closer, and then merge at the location of the
critical line, and only one image with the same parity as the original
source parity remains when the source moves outside of the caustics.
If a cusp caustic lies outside the radial caustic as described in 
Fig. 3, such a cusp is termed ``naked cusp." 
If a source lies close to a naked cusp, one or three highly 
magnified images appear without any additional images. 
The first discovered giant arc in A370 is likely to be a highly
magnified image due to a naked cusp (Fig. \ref{fig:a370}).

For simple gravitational lenses (with one main clump) we can
distinguish between two types of caustics.  Tangential caustics occur when the
deformation of the merging images is tangential (nearly parallel to
the critical line).  Radial caustics occur when the deformation of the
merging images is radial (nearly perpendicular to the critical line).

\begin{figure}
\vspace{20cm}  
\special{epsfile=fig3a.ps rotation=-90 hoffset=180 voffset=575 hscale=0.8 vscale=0.8}
\special{epsfile=fig3b.ps rotation=-90  hoffset=400 voffset=520 hscale=0.5 vscale=0.5}
\special{epsfile=fig3c.ps rotation=-90  hoffset=155 voffset=360 hscale=0.8 vscale=0.8}
\special{epsfile=fig3d.ps rotation=-90 hoffset=390 voffset=360 hscale=0.5 vscale=0.5}
\special{epsfile=fig3e.ps rotation=0  hoffset=20 voffset=130 hscale=0.3 vscale=0.3}
\special{epsfile=fig3f.ps rotation=0 hoffset=220 voffset=160 hscale=0.3 vscale=0.3}
\caption{The caustics  and  critical lines of an elliptical lens.  
Left and right panels describe the caustics in the source plane
and the critical lines in the image plane, respectively. 
Four cusp caustics appear at the corners of diamond caustics. 
The other parts of the caustics are folds. 
The appearance of the images
corresponding to each source position are also shown.
The minor axis of the elliptical lens potential is the vertical direction.
The ellipticity of the lens potential is larger in the case
shown in bottom figures than in the others. 
Naked cusps appear.  When a source lies on a naked cusp,
one highly magnified merged image appears without any additional 
images.} 
\label{fig:3}
\end{figure}

Detection of a  radial arc introduces a strong constraint on the degree of the
central mass concentration in the cluster.  For example, in the case
of a non-singular mass distribution (finite core) we expect a radially
stretched image. This is clearly seen in the case of a circular
symmetric lens (Fig. 2).  When the source approaches the radial
caustics, two images approach each other  across a radial critical line
along the radial direction and merge into a radially stretched single
image.  If the density distribution of the lensing object in the
central region can be approximated by $\rho\propto r^{-\gamma}$ and
$\gamma < 2$, the gravitational potential and the refractive index are
finite at the centre of the lens. Hence, the gravitational delay is
finite, and a light ray can find its way through the centre of the
lens.  On the other hand, if $\gamma\geq 2$, the gravitational
potential is singular ( e.g. $\phi(\vec{r}) \rightarrow -\infty$ when
$|\vec{r}| \rightarrow 0$) and the gravitational delay becomes infinite
at the centre. Therefore, a light ray which passes near the centre never
arrives to the observer.  For example, a point mass lens and a
singular isothermal sphere do not have radial caustics and will
produce two multiple image.\cite{Blandford87}\tocite{Mellier98}  On
the other hand, the NFW density profile may have a radial
arc\cite{BART96} since it has $\gamma=1$ in the central region.  It
has been sometimes said that the detection of a radial arc is 
evidence of the finite core radius of the cluster mass
distribution.\cite{Fort94}  However, the only conclusion that can be
drawn from the detection of the radial arc is that the density
profile of the cluster central region is flatter than $\rho\propto
r^{-2}$, e.g.  $\gamma < 2$.

\begin{figure}
\vspace{9cm}  
\special{epsfile=fig4a.ps  hoffset=3 voffset=250 hscale=0.6 vscale=0.6}
\special{epsfile=fig4b.ps  hoffset=222 voffset=250 hscale=0.6 vscale=0.6}
\caption{The time delay surface for a circular lens with $\beta=0$ (left panel)
and $\beta\ne 0$ (right panel), where the source position 
is shifted to negative $y$-direction.  A contour line corresponding to  
the same height as a saddle 
point of the time delay surface is also drawn.  Each stationary point is named 
according to their second derivatives: {\rm H}(high), {\rm S}(saddle) 
and {\rm L}(low) denote the local maximum, the saddle and the local minimum points,
respectively.  The isochrone appearing in the right panel is named ``lima\c{c}on".
 }
\label{fig:4}
\end{figure}

\begin{figure}
\vspace{20cm}  
\special{epsfile=fig5a.ps  hoffset=5 voffset=500 hscale=0.7 vscale=0.7}
\special{epsfile=fig5b.ps  hoffset=102 voffset=520 hscale=0.7 vscale=0.7}
\special{epsfile=fig5c.ps  hoffset=100 voffset=250 hscale=0.7 vscale=0.7}
\caption{The time delay surface for an elliptical lens with $\vec{\beta}=\vec{0}$
(top left panel)
and  $\beta_x>0\;(\beta_y=0)$ (top right and bottom panels).
Contour lines corresponding to the same height as a saddle 
point of the time delay surface and the name of each stationary point
are also indicated.  In the top two panels, two saddle points appear. 
Shifting the source position further from the top right panel, 
two Ls and one S merge into one L, and one obtains the case described by bottom panel. 
The top right panel has both a lima\c{c}on and a lemniscate as isochrones. 
 }
\label{fig:5}
\end{figure}

The time-delay surface helps us to visualize image configurations for the
complex lens model. As we can see from Eq. (\ref{eq:11}), the geometrical
delay surface is a concave paraboloid, and the gravitational delay 
adds a bump to the paraboloid.  For a circular symmetric lens with a
finite core radius and a perfect alignment source-lens-observer
(Fig. 4), $\beta=0$, and the time-delay surface will display one
maximum (H) and one circle as minimum (L).  The circle L corresponds
to the Einstein ring.  For small but finite values of $\beta$, the Einstein
ring disrupts into one saddle point (S) and one minimum.  The total
number of the stationary points (hence the number of images) is 3. The
total parity of each stationary point is even for H and L images,
and odd for S image.  
If we increase $\beta$,
H and S will merge (at the critical line) along the radial direction
forming the radial arc, and then disappear, and only the L image remains.
Note that if the central potential depth is infinite, there is no maximum H
and no radial arc can exist.  For an elliptical potential and in the
perfect alignment case, the ring L disrupts into two Ss (along the major
axis of the potential) and two Ls (along the minor axis of the
potential), because the gradient of the effective lensing potential is
different along major and minor axes (Fig. 5).  First we consider the
case in which a cusp lies inside of the radial caustic.  If we shift the
source position from perfect alignment along the major axis, two
Ls and one S will move closer along a curved trajectory tangential to
the lens potential contour and merge, forming a tangential arc.  The
position of the source at this stage lies at the cusp caustics.
If we further shift the source position, the merged image converges
into a single L image, and H and S approach along the radial direction and
merge into one radial arc, where the source is on the radial caustics.
Further outward shifting of the source position causes H and S to disappear,
and only the L image remains.

In the case of a naked cusp, first H and S
approach along the radial direction and merge into one radial arc, where
the source is on the radial caustic, and finally they disappear if we shift
the source position from the perfect alignment along the major axis.
If we further shift the source position, two Ls and one S become
closer and merge, forming a tangential arc.  The position of the
source at this stage lies at the naked cusp caustics.  The elliptical
lens can have five or three or one images, and the total number of images
always changes by $\pm2$ when the source crosses a caustic.

In the case of an  elliptical lens, one can see another type of
important caustic for cluster lensing, that is lips caustics (Fig. 6).
We start with the perfect alignment case (Fig. 5).  As decreasing the
depth of the effective lensing potential, a gradient along the major
axis is getting flatter.  Finally, two Ss disappear and a central H
changes into S.  Such caustics are called `lips caustics' because of their
shape.\cite{SEF92}  The time delay surface and shape of the
lips caustics are drawn in Fig. 6. If an extended source fills in the
lips caustics, one S and two L images merge into one straight arc.  If
we further decrease the depth of the effective lensing potential, one
S and two Ls merge and disappear, and only one L remains.  Since the
Laplacian of the time-delay potential reads
\begin{equation}
\Delta t(\vec{\theta})\propto
2-\Delta\psi(\vec{\theta})=2(1-\kappa(\vec{\theta})), \label{eq:12}
\end{equation}
the condition of the maximum point, $\Delta t(\vec{\theta})<0$, reads 
$\kappa(\vec{\theta})>1$.  Therefore, to have an H image, the surface mass
density of the lens must be super critical, and the lens must be super
critical for the circular symmetric lens to be able to have strong
lensing events and for the elliptical lens to have five images.
However, in the case of lips caustics, there is no H, and the lens does
not have to be super critical.  These diagnostics of the lips shows
that the lips caustics appear when the surface mass density of the
lens is marginally critical.

\begin{figure}
\vspace{10cm}  
\special{epsfile=fig6a.ps  hoffset=140 voffset=285 hscale=0.5 vscale=0.5}
\special{epsfile=fig6b.ps  hoffset=10 voffset=110 hscale=0.5 vscale=0.5}
\special{epsfile=fig6c.ps  hoffset=220 voffset=130 hscale=0.5 vscale=0.5}
\caption{The lips caustic.  The time delay surface for the marginally critical elliptical
lens with $\beta=0$ is shown in the top panel. The major axis of the elliptical lens
potential is the $x$-direction.
The bottom left and right panels show the caustic and critical line in this case,
respectively.  This caustic is called a ``lips caustic" because of its shape. 
The appearance of the images
corresponding to each source position are also shown.
If the central source size is larger and touches the fold caustic line,
three multiply lensed images  merge into one straight arc.
Even if the source appears on the cusp, the curvature radius
of the arc is large, and a nearly straight arc is obtained.}
\label{fig:6}
\end{figure}

\begin{figure}
\vspace{17cm}  
\special{epsfile=fig7a.ps  hoffset=5 voffset=482 hscale=0.4 vscale=0.4}
\special{epsfile=fig7b.ps   hoffset=200 voffset=490 hscale=0.4 vscale=0.4}
\special{epsfile=fig7c.ps  hoffset=5 voffset=370 hscale=0.4 vscale=0.4}
\special{epsfile=fig7d.ps  hoffset=190 voffset=400 hscale=0.4 vscale=0.4}
\special{epsfile=fig7e.ps hoffset=10 voffset=235 hscale=0.6 vscale=0.6}
\special{epsfile=fig7f.ps hoffset=230 voffset=235 hscale=0.6 vscale=0.6}
\caption{The beak-to-beak caustic. 
Two critical circular lenses are put on the same lens plane with a
small separation.  The left top and left middle panels are caustics, 
and the right ones are critical lines, respectively.  The strengths of the lenses are 
weaker; that is, the source is closer to the lens in the case shown in the top panels.  
As the source gets closer to the lens, two fold caustic lines appearing in
the central region merge and finally disappear (changes from the middle panel 
to the top panel). 
The caustics shown in the top panel are  called ``beak-to-beak" caustics,
owing to their shape.
The appearence of the images
corresponding to each source position are also shown.
A highly stretched straight arc is obtained when a source exists at the 
middle of two lenses.  
The time delay surfaces and their isochrones for the lens system corresponding to the middle 
panel are shown in the bottom panels. The left panel is shown  for the case 
when the source,  the observer and the middle point of the two lenses are aligned. 
The case in which  the  source position is slightly shifted toward the centre 
of one of the lenses is shown in the right panel. 
 }
\label{fig:7}
\end{figure}

The beak-to-beak catastrophe is another example of a marginally critical
lens.\cite{SEF92}  It is formed of two marginally critical lenses
close to each other in the same lens plane.  In Fig. 7 the behavior of
time delay surfaces for the observer at the middle of two circular
lenses in the case that the strengths of two equal lenses are similar, is shown.
Decreasing the strength of the lens (that is, decreasing the absolute
value of the amplitude of the effective lensing potential) corresponds
to putting the source at lower redshift for a fixed lens mass distribution
and fixed cosmological parameters.  In the case shown in Fig. 7, two
L images and one S image appear along the line perpendicular to the line
across the two lenses when the source is placed in the middle of the two lenses.
When decreasing the strength of the lenses, two Ls and one S approach
and merges, and only one L image remains.  The time-delay surface and
caustic for this case are shown in Fig. 7.  These are termed
a `beak-to-beak caustic.'  Both lips caustics and beak-to-beak caustics can form
straight arcs.\cite{Pello91,MATH92}  Lips and beak-to-beaks are not
listed as elementary catastrophes.  The reason for this is that these are the
cross sections by the source plane of the cusp ridges associated with
any of the elementary catastrophes in a control space of
greater than three dimensions.
This situation is illustrated well by Fig. 6.12 in
Schneider, Ehlers and Falco.\cite{SEF92}
 
As mentioned at the beginning of this section, three more higher
order elementary catastrophes appear, but they always are a combination
of cups and folds. Readers can consult the textbook by Schneider,
Ehlers  and Falco\cite{SEF92} with regard to the application of the higher order
catastrophe for the lensing.

\subsubsection{The strong lensing cross section}

For  statistical studies of  strong lensing events, it is
important to compute the lensing cross-section: the probability to
observe the image with a magnification larger than a threshold value $\mu$.
The computation of the cross-section near fold
caustics\cite{Blandford87} gives a good first order approximation of
the total cross section.  Indeed, most of caustics are folds, and the
contribution to the total cross section from a cusp is generally
negligible, since they are points.  Without loss of generality, the
origin of the coordinate system in the source plane can be put on a
fold caustic, and the direction of the $x$ and $y$ coordinates can be defined as
tangential and perpendicular to the fold caustic, respectively.  As
long as we are interested in a local estimate, the $x$-coordinate can be
approximated as the caustic line, and the magnification then only
depends on the $y$-coordinate, the distance to the caustic line.

Thus, the cross section near the fold caustics is given by
\begin{equation}
\sigma(\ge \mu)=\oint\int_0^{\beta_y(\mu)}D_{\rm s}^2 d\beta_x\;d\beta_y,
\label{eq:13}
\end{equation}
where $\beta_y(\mu)$ is the position where the magnification
reaches $\mu$ and the integral over $\beta_x$ is taken along the
fold caustic.  This equation can be easily rewritten as
\begin{equation}
\sigma(\ge \mu)=\oint\int_0^{\theta_y(\mu)}D_{\rm s}^2
H(\vec{\theta})d\theta_x\;d\theta_y,
\label{eq:14}
\end{equation}
where we have used the fact that the caustic line is mapped onto the
critical line. To avoid double counting the area, we integrate only on
one side of the critical line, although two images with approximately
the same magnification appear in both sides of the critical lines.
Since the magnification is infinite at the critical line, the Hessian
$H(\vec{0})$ is zero on the critical line. Thus the derivatives of the
Hessian along the critical line are also null, in particular,
$\partial H(\vec{0})/\partial \theta_x =0$.  Therefore, the
Taylor expansion of the Hessian around the critical line is
\begin{equation}
H(\vec{\theta})={\partial H(\vec{0})\over \partial \theta_y}
\theta_y.
\label{eq:15}
\end{equation}
Then $\theta_y(\mu)$ is defined by $\mu$ as
\begin{equation}
\mu=H(\theta_y(\mu))^{-1}=\left(\left\vert{\partial H(\vec{0})\over \partial 
\theta_{\rm y}}\theta_y(\mu)\right\vert\right)^{-1}.
\label{eq:16}
\end{equation}
Finally we obtain
\begin{equation}
\sigma(\ge \mu)={1\over 2\mu^2}D_{\rm s}^2\oint {d\theta_{\parallel}\over
\vert\nabla_{\perp} H\vert_{H=0}},
\label{eq:17}
\end{equation}
where $d\theta_{\parallel}$ is the length element along the critical line and
$\vert\nabla_{\perp} H\vert_{H=0}$ is the gradient of the Hessian in the
direction of the perpendicular to the critical line evaluated on the
critical line, and to first order it does not depend on $\mu$.  We
thus conclude that the total cross section to observe lensed images
with magnification larger than $\mu$ is proportional to $\mu^{-2}$.

\section{Constraining the mass distribution of clusters}

\subsection{Modelling the mass distribution with strong lensing}

\begin{wrapfigure}{r}{6.6cm}
\vspace{8cm}  
\special{epsfile=fig1.ps  hoffset=-10 voffset=230 hscale=0.4 vscale=0.4}
\caption{The {\sl HST}/WFPC2 image of CL0024 +1654 at $z=0.39$. One can clearly see 
that one galaxy is multiply imaged, 7 images in this case. The change of the 
parity and blue nature of the lensed images can be observed.}
\label{fig:cl0024}
\end{wrapfigure}

\begin{figure}
\vspace{20cm}  
\special{epsfile=fig8a.ps  rotation=-16 hoffset=130 voffset=580 hscale=0.5 vscale=0.5}
\special{epsfile=fig8b.ps  rotation=-90 hoffset=360 voffset=270 hscale=0.5 vscale=0.5}
\caption{The central part of A370 taken  by  {\sl HST}/WFPC2 
image in F675W the top panel.\cite{Bez98}  Overlaid is the mass distribution (black thin lines). 
The most important arclets and multiple images
are  colored. 
The radial arc can be observed in the south clump. 
The contour map of the X-ray emission obtained by {\sl ROSAT}/HRI is 
shown in the bottom panel.  This is taken from Hashimotodani.\cite{Hashi99} 
The crosses indicate  the positions of 
the two cD galaxies.  There is a small cave close to the northern cD. 
North is up in both figures.}
\label{fig:a370}
\end{figure}

Strong lensing events, such as GLAs and multiply imaged arcs
(out-standing example is shown in Fig. \ref{fig:cl0024}), constitute
a powerful tool to constrain the mass distribution of cluster cores.
As we have shown in the previous section, 
the simplest way of estimating the mass of a
cluster with strong lensing events is assuming circular symmetry for
the lens and to use Eq. (\ref{eq:8}), where the Einstein radius 
$\theta_{\rm E}$ is estimated from the position or curvature of the arc 
or from the distance between multiple images. Such an estimate is, however, 
{\it very rough} and can easily over or underestimate the effective
projected mass in the core.
 
More accurate modelling of the cluster mass distributions are based on
multiple images. The idea here is to optimize the cluster mass model
in order to reproduce the observed sets of multiple images within the
observational uncertainty.  A $\chi^2$ can be defined by the quadratic
sum of the difference of the central positions of each image mapped
onto the source plane divided by the observational uncertainty scaled by
the lens mapping.  Alternatively, it can be defined as the quadratic sum of
the difference between the actual observed images and the images of a
fiducial source.  The elongation and orientation of each image can in
principle be added to the optimization procedure.  However, due to the
relatively large observational uncertainty compared to the image
position, they generally impose a loose constraint.  Since the number of
available observational constraints is not large, the number of
parameters describing the cluster mass distribution should be of the
same order as the number of constraints.  For this reason, the mass
model is often described by a linear sum of mass clumps described
analytically.  One of the first best known examples of a cluster mass
model is the model of A370 by Kneib, Mellier, Fort and
Mathez.\cite{KMFM93} Their model is very simple in the sense that it has
only two mass clumps centred on the two giant ellipticals of that
cluster, and was constrained by the giant arc (a triple image) and an
other triple image, B2-B3-B4.  From their model they could predict the
redshift for the source of B2 to be $z\sim 0.865$.  Figure
\ref{fig:a370} shows the central part of A370 seen with {\sl HST}/WFPC2 in F675W,
with the improved mass distribution model by B\'ezecourt et
al.\cite{Bez98} that extends the former model including cluster
galaxies (following the prescriptions of Kneib et al.\cite{Kneib96}
The latest model of A370 is based on two sets of double images and four
sets of triple images with a spectroscopic redshift for two of them.  The
fact that the source redshift of B2 was first predicted by Kneib
et al.\cite{KMFM93} and then confirmed spectroscopically by
B\'ezecourt et al.,\cite{Bez98} strengthens support for the reliability of the
model.

Cluster-lenses mass models have been considerably improved with the
refurbishment of the Hubble Space Telescope (HST). Indeed, the high
image-resolution of the {\it HST}/WFPC2 camera on a reasonably wide
field makes it a unique instrument to probe the mass distribution of
cluster cores.  As was first shown by Kneib et al.,\cite{Kneib96}
thanks to the identification of a larger number of multiple images, it
is possible to constrain the mass distribution on the scale from $\sim$500
kpc down to the galaxy scale $\sim$10 kpc. Strong lensing events {\it
  do} constrain the mass distribution of the cluster core, and more
specially the mass-to-light ratio of cluster galaxies. Furthermore, by
measuring the deviation of the shear field from the prediction of the
cluster potential which is constrained by strong lensing events, the
amplitude of the perturbations due to the cluster member galaxies can
be evaluated.\cite{NAT98} Therefore, the mass-to-light ratio of the 
cluster member galaxies can be constrained. Such an approach combined
with the weak (galaxy-galaxy) lensing approach can measure even more
accurately the masses of galaxies.\cite{NAT98,Geiger98}

\subsection{Comparison with X-ray results}

Clusters of galaxies contain huge amounts of hot gas, which emit X-ray
through bremsstrahlung. This hot gas constitutes the intra-cluster medium
(ICM).  Since the sound crossing time of the ICM through the cluster core
is $\sim 10^8$yr and generally is much less than the age of the
cluster ($\sim {\rm several}\times 10^9$yr), the ICM could be 
described well by hydrostatic equilibrium.  Therefore, if the gas density
distribution and the hot gas temperature distribution are measured,
the mass distribution of the cluster can  be obtained by solving the
equation of hydrostatic equilibrium.  The {\sl
  ROSAT}/HRI\cite{Truem84} had a good spatial resolution in 0.1--2keV
and has provided precise information on the distribution of the ICM.
The {\sl ASCA}\cite{Tanaka94} satellite has a good spectral resolution
in the 0.5--10keV band and is the first instrument able to
measure the ICM temperature in the distant lensing cluster (despite a poor
spatial resolution).  We were very lucky to have both complementary
X-ray satellite past five years, while the lensing study of the cluster
mass distribution was emerging as a new powerful tool to constrain
cluster mass distributions.  Lensing and X-ray studies of the cluster
mass distributions are complementary.  While X-ray studies of 
cluster mass distributions require some assumptions on the thermal and
dynamical state of the ICM, lensing directly maps the surface mass
density of the cluster, regardless of its dynamical and thermal state.
Therefore, lensing studies of cluster mass distributions are more
reliable than X-rays {\it if} sufficiently accurate lensing constraints are
available.  Comparison of the two results provides information on the
thermal and dynamical states of clusters.  Such a comparison may also
shed new light on the formation and evolution of clusters which
tightly links them to modern cosmological interests.

Peebles\cite{Peebles93} wrote in 1993, {\it ``[...]  we are going to
  extrapolate the physics that is known to be successful until it is
  seen to fail. [...] the extrapolation out in space and back in time
  is by no means without empirical support. [...]  gravitational
  lensing of background galaxies by the mass concentrations in
  clusters of galaxies, as analyzed in general relativity theory, is
  consistent with the masses derived from the motions of the galaxies
  and from the plasma pressure within the clusters.  The relevant
  length scale here --- the impact parameter at the center --- is ten
  orders of magnitude larger than that of the precision tests of
  general relativity in the Solar System and in binary pulsar systems,
  a remarkable extrapolation."}
\begin{wrapfigure}{r}{6.6cm}
\vspace{9.2cm}  
\special{epsfile=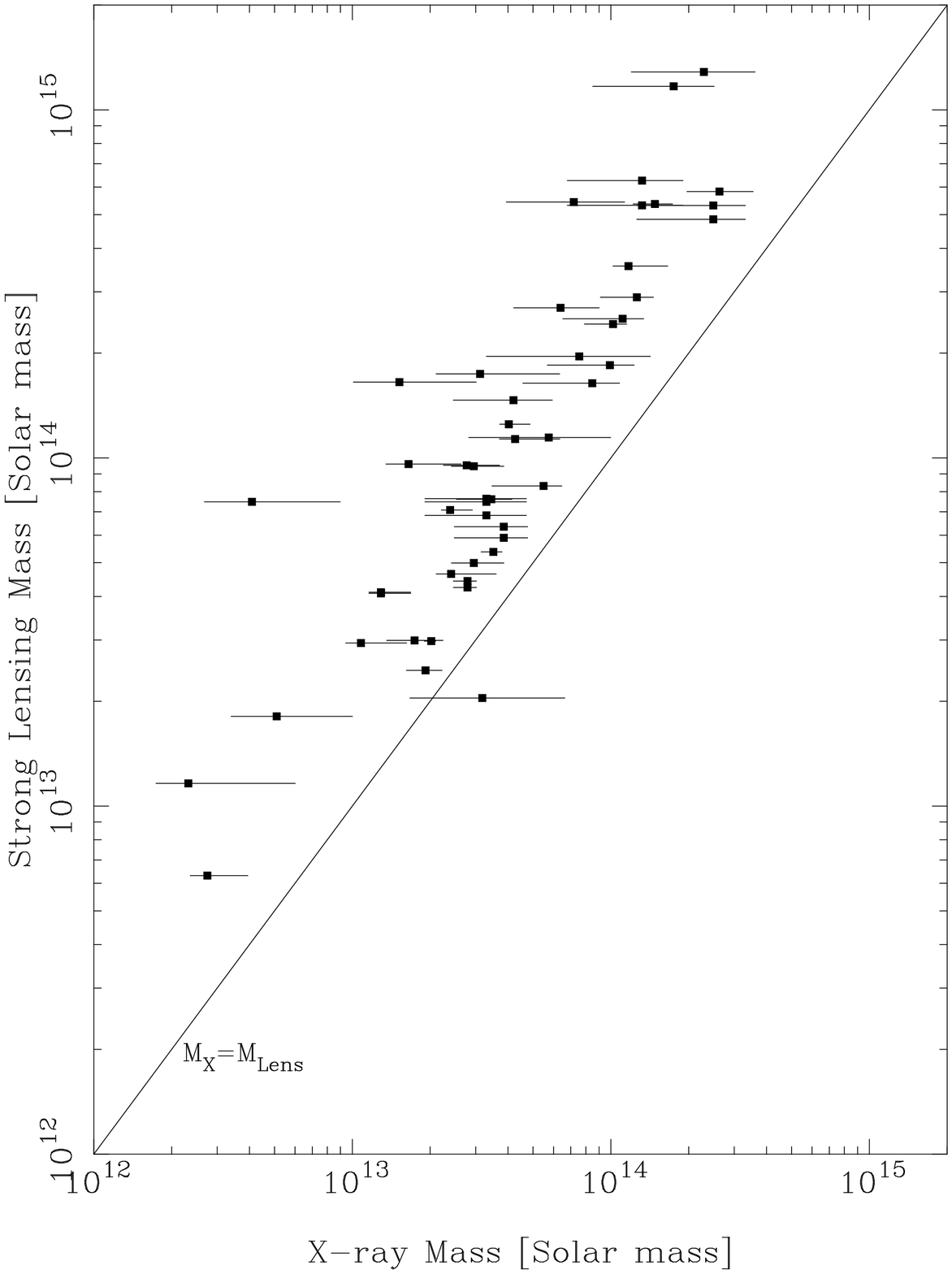  hoffset=-10 voffset=300 hscale=0.4 vscale=0.4}
\caption{The comparison of the cluster mass obtained by X-ray observations
with those obtained by modelling of strong lensing events.  The lensing masses 
are quoted  from the literature, mostly from Wu and Fang.\cite{WU97} 
This figure is quoted from Hashimotodani.\cite{Hashi99} }
\label{fig:XvsL}
\end{wrapfigure}
However, soon after the publication of Peebles' textbook, it was
claimed that the mass in the cluster core region estimated from 
strong lensing effects is systematically a few times larger than the
mass estimated by X-ray observations under the assumption of the
hydrostatic equilibrium of the ICM.\cite{Loeb94}\tocite{WU98}  The
simple comparison discussed in these papers has created a great deal of
excitement and discussion, in order to try to explain such
discrepancy.  Hashimotodani\cite{Hashi99} showed that the discrepancy
between the lensing and X-ray approaches seems to be systematic on a
large sample of 30 distant cluster lenses (Fig. \ref{fig:XvsL}).
Such analyses on such large samples was first made by Wu and
Fang,\cite{WU97} but their X-ray mass estimates were not an accurate, since
they were not using {\sl ROSAT}/HRI and {\sl ASCA} data.
Allen\cite{Allen98} also analyzed the {\sl ROSAT}/HRI and {\sl ASCA} data
of 13 distant cluster lenses and obtained the first reliable mass
estimations based on X-ray observations, and has made the first robust
confirmation of the existence of the systematic discrepancy.


\subsection{Understanding the mass discrepancy}

In this section we introduce ideas proposed to solve
the mass discrepancy. We also attempt to discuss the best current understanding
of the problem.

\subsubsection{Non-thermal pressure}

Loeb and Mao\cite{Loeb94} proposed that non-thermal pressure induced
by equipartition turbulent and magnetic pressure plays a crucial role
in supporting the ICM against the cluster gravitational pull.
Therefore, if one neglects the contribution of these non-thermal
pressures, one could easily underestimate the gravitational mass of
the cluster from the ICM distribution by a factor of 2--3.  They
concluded that the observed high Faraday rotation measures (hereafter RM)
in the directions of radio lobes in several (low-$z$) clusters would be consistent with
the required equipartition magnetic field needed in lensing clusters to
resolve the discrepancy.  However, Makino\cite{Maki97} pointed out that
the magnetic field strength estimated from the RM in the direction of
background radio sources is an order of magnitude less than the
equipartition magnetic field strength. He argued that the extremely
high RM obtained in the direction of the radio lobe is likely to be intrinsic to
the radio lobe but not to the ICM. The magnetic field strength in 
cluster lenses is undetermined at present. It is, however, constrained
by hard X-ray and radio observations. Indeed, extended radio 
halo caused by  synchrotron radiation 
has been detected from several nearby clusters.\cite{Kim89}
The relativistic electrons in the cluster, which constitute the source 
of the synchrotron radiation, produce a hard X-ray tail in 
the cluster X-ray spectrum by inverse Compton scattering of
the cosmic microwave background radiation.\cite{Sarazin86}
It is
expected that the next generation of X-ray satellites ({\it Astro-E}) will judge
the possibility of the equipartition magnetic field in the ICM by
measuring the hard X-ray tail.

\subsubsection{Elongation of the cluster and 3D mass distribution}

Loeb and Mao\cite{Loeb94} pointed out that the cluster elongation along
the line of sight is not a plausible explanation since it would
require a large axial ratio that would cause a dynamical
instability.\cite{MH91}  Furthermore, it is very unlikely that all the
lensing clusters are elongated along the line of sight by accident.
However, the possibility of the superposition of several clusters along
the line of sight could easily explain the discrepancy observed for some 
optically selected lensing cluster.  For example, the deep wide field
redshift survey of Cl0024+1654\cite{Soucail99,Czoske99} seems to favor
the existence of the superposition of two close-by clusters.

\subsubsection{Cooling flow}

Clusters of galaxies are categorized into two groups according to the nature of
the distribution of the ICM in the cluster core. The clusters in which
the X-ray emission has a prominent central peak that coincides with
the center of a bright elliptical galaxy is called {\it XD}. 
Clusters in which the X-ray emission is dispersed and does not exhibit a
clear coincidence between the X-ray peak and the bright galaxy center is
called {\it non-XD}.\cite{Forman88} The XD clusters are believed to
harbour a cooling flow.\cite{Fabian94}

The multi-phase nature of the ICM in the central part of clusters due
to cooling flow has been considered as a possible source of the
underestimation of the X-ray cluster mass.\cite{Allen98,AFK96}
Allen\cite{Allen98} studied X-ray data of 13 lensing clusters of
galaxies and broke his sample into three groups, on their
central cooling time: cooling flows, non-cooling flows, and
intermediate samples. He showed that the mass discrepancy is
significantly reduced for cooling flow samples by applying the cooling
flow model to the spectral fitting.  If the cooler components induced
by cooling flow exists, the temperature of the hottest component should
become higher by a factor of 2--3 than the temperature obtained by a
single phase model fitting of the observed emissivity-weighted
temperature. Since the hydrostatic mass depends linearly on the temperature
of the hottest phase, it increases the X-ray mass. He also noticed that
the centre of the X-ray emission is significantly shifted from the
centres of curvature of the lensed arc for the non-cooling flow
sample.  Therefore, these clusters may not be in dynamical
equilibrium and probably are post-mergers.  However, several arguments
do not support this idea.  {\sl ASCA} data of nearby XD clusters are
fitted well by a two-temperature phase model in the core and with a
single phase model in the outer region.\cite{Ikebe96} The hotter phase
stays isothermal throughout the cluster, and the weighted averaged
temperature throughout the cluster is not so much different from the
temperature of the hotter phase; this seems to be contradictory
to Allen's\cite{Allen98} analysis.  The distant cluster RXJ1347-1145
is the most X-ray luminous cluster ever discovered and is thought to
have an extremely massive cooling flow, according to Allen.\cite{Allen98}
Recently a radial profile of the Sunyaev-Zel'dovich (hereafter SZ)
increment was measured by {\sl JCMT}/SCUBA\cite{Kom98} for this
cluster.  It was shown that the combined fitting\cite{HAT99} to the X-ray
spectrum obtained by {\sl ASCA} and to the SZ radial profile puts a
strong constraint on the strength of the possible cooling flow and
leads us to reject the massive cooling flow model proposed by
Allen.\cite{Allen98}

Clearly, deeper insight here is needed to improve our knowledge on the
cooling flow model. The next generation of X-ray satellites, like {\sl
  AXAF/XMM} with their spatially high-resolution temperature
measurement (of cluster-lenses) and {\sl Astro-E} with its high
resolution X-ray line spectroscopy (of nearby cooling flow candidate
clusters), and on-going (e.g. {\sl JCMT}/SCUBA) and future projects 
(e.g. {\sl LSA/MMA, LMSA}) of mm and submm observations of SZ and 
dust emission from clusters,\cite{Edge99}
will soon help us in understanding the physical properties
of the ICM in a cooling flow cluster.

\subsubsection{ICM not in dynamical equilibrium}

Allen\cite{Allen98} also suggests that for non-cooling flow cluster,
the main reason for the discrepancy is that these clusters are not 
in dynamical equilibrium, as they are post-mergers.
The evidence given to support this idea (see also Hashimotodani\cite{Hashi99})
is that the centre of the X-ray emission is significantly shifted from the 
centres of curvature of the lensed arc.  
Such a possibility was first proposed by Kneib et al.,\cite{Kneib95}
in their detailed mass distribution model of A2218 based on lensing
data and compared with the {\sl ROSAT/}HRI X-ray map.
The lensing model predicts a bimodal distribution
centered  on the two brightest cluster galaxies.
The distribution of the X-ray emitting gas is, however, not consistent
with the lensing mass model: the X-ray bright core is elongated perpendicular
to the elongation of the lens model, with a shift in the position.
Moreover, there is no secondary peak in the X-ray emission associated with
the secondary clump inferred from the lens model.
Thus, they suggested that the ICM in A2218 is highly disturbed
by the passage of a subcluster, and the ICM is no longer in hydrostatic 
equilibrium in the cluster core.

This possibility is also supported by other evidence.  The left-hand
side of Eq. (\ref{eq:sig0}) is $(830{\rm km/sec})^2=6.9\times
10^{15}{\rm cm^2/sec^2}$ for the main clump model presented in Kneib et
al.\cite{Kneib95}  The best fitting values to the X-ray data yields
$(\beta/0.62)(k_{\rm
  B}T/7.04{\rm keV})$ $=6.6\times 10^{15}{\rm cm^2/sec^2}$  
for the right-hand side of Eq. (\ref{eq:sig0}).  
These two
values are consistent within X-ray measurement errors for temperature
$7.04\pm 0.07$keV\cite{MS97} and for slope parameter of X-ray surface brightness
$\beta=0.62^{+0.13}_{-0.09}$\cite{Hashi99} where errors are 90\% one parameter
errors.  The main discrepancy in this cluster can be explained by the
large X-ray core radius of $53''$,\cite{Birk94,Maki98} compared to the
value in the lensing model, $\sim 20''$.  A simple check of this can be done in
the following way.  Inserting $\theta_c=53''$, $\theta_{\rm E}=23''$ and
$z_s=2.6$ into Eq. (\ref{eq:kT}) yields $k_{\rm B}T=17$keV, while
$\theta_c=20''$ yields $k_{\rm B}T=9.5$keV, which is close to the
measured temperature.  If the post-merger causes the cluster central
gas to disperse and to thus forces it away from equilibrium, but allows for the dark matter
distribution to remain centrally concentrated (as suggested by lensing), one
should use a core radius of $\sim 50-100h_{50}^{-1}$kpc (about $20''$)
for the dark matter distribution (which is the typical X-ray core
radius for an XD cluster).  Then, assuming that the gas distribution at
the cluster outskirts is in hydrostatic equilibrium, where the observed
$\beta$ value and the temperature are measured, the mass discrepancy
problem for A2218 is solved.  The last assumption may be safely
applied, since the mass of the subclump supposed to have collided is
much smaller than the mass of the main clump.\cite{Kneib95}

\subsubsection{Effects of complex mass distribution}

Recent studies have revealed other plausible
possibilities to reconcile the discrepancy.  The main
reason here is overestimation in the lensing mass
determinations.
By definition, a large number of galaxies populates the cluster core.
Therefore, we cannot neglect the contribution of their stellar mass
and their dark halos. With such complex models, the total mass needed 
to produce arcs is significantly decreased compared to smoothed and
symmetric models.

\begin{figure}
\vspace{8cm}  
\special{epsfile=a2390ic.ps   hoffset=10 voffset=250 hscale=0.7 vscale=0.7}
\caption{The central part of cluster A2390 seen with {\sl HST}/WFPC2/F814W. 
A-B-C is a straight arc. It is in fact composed of two objects: B-C at $z=0.913$ and A at 
$z=1.03$. The objects H3a-b and H5a-b are two sets of multiple images with 
$z=4.06$.\cite{Pello98}}
\label{fig:a2390}
\end{figure}

For example, Abell 2390 has a ``straight'' arc.\cite{Pello91} It
requires a special geometry for the mass distribution of the lensing
object and imposes a strong constraint on the cluster mass distribution
(Fig. \ref{fig:a2390}).  Pierre et al.\cite{Pier96} presented a
detailed analysis of the {\sl ROSAT}/HRI image and found an extended
local sub-peak of X-ray emission which might be associated with the existence
of a dark matter clump.  This clump would act as a perturber, just as
predicted by Pell\'o et al.,\cite{Pello91} to reproduce the straight
arc. In fact Pierre et al.\cite{Pier96} successfully constructed 
a lens model of A2390 defined by two mass clumps. B\"{o}hringer et
al.\cite{BOEH98} measured the temperature of this cluster and studied
its X-ray data, which exhibits a very smooth, centrally
peaked X-ray emission, and evidence of massive undisturbed cooling
flow, indicating that the ICM is in hydrostatic equilibrium in the
cluster potential. They showed that the X-ray mass estimation assuming
that the cooling flow has little effect on the temperature measurement
agrees well with the lensing mass inferred by Pierre et
al.\cite{Pier96} If the lensing mass of this cluster is estimated
with the formula for the circular symmetric lens, the estimated mass
is several times larger than the X-ray mass.  These results indicate
that the introduction of a small asymmetry in the mass distribution of
the cluster core can substantially reduce the lensing mass compared to
the circular symmetric approximation.

\begin{wrapfigure}{r}{6.6cm}
\vspace{8cm}  
\special{epsfile=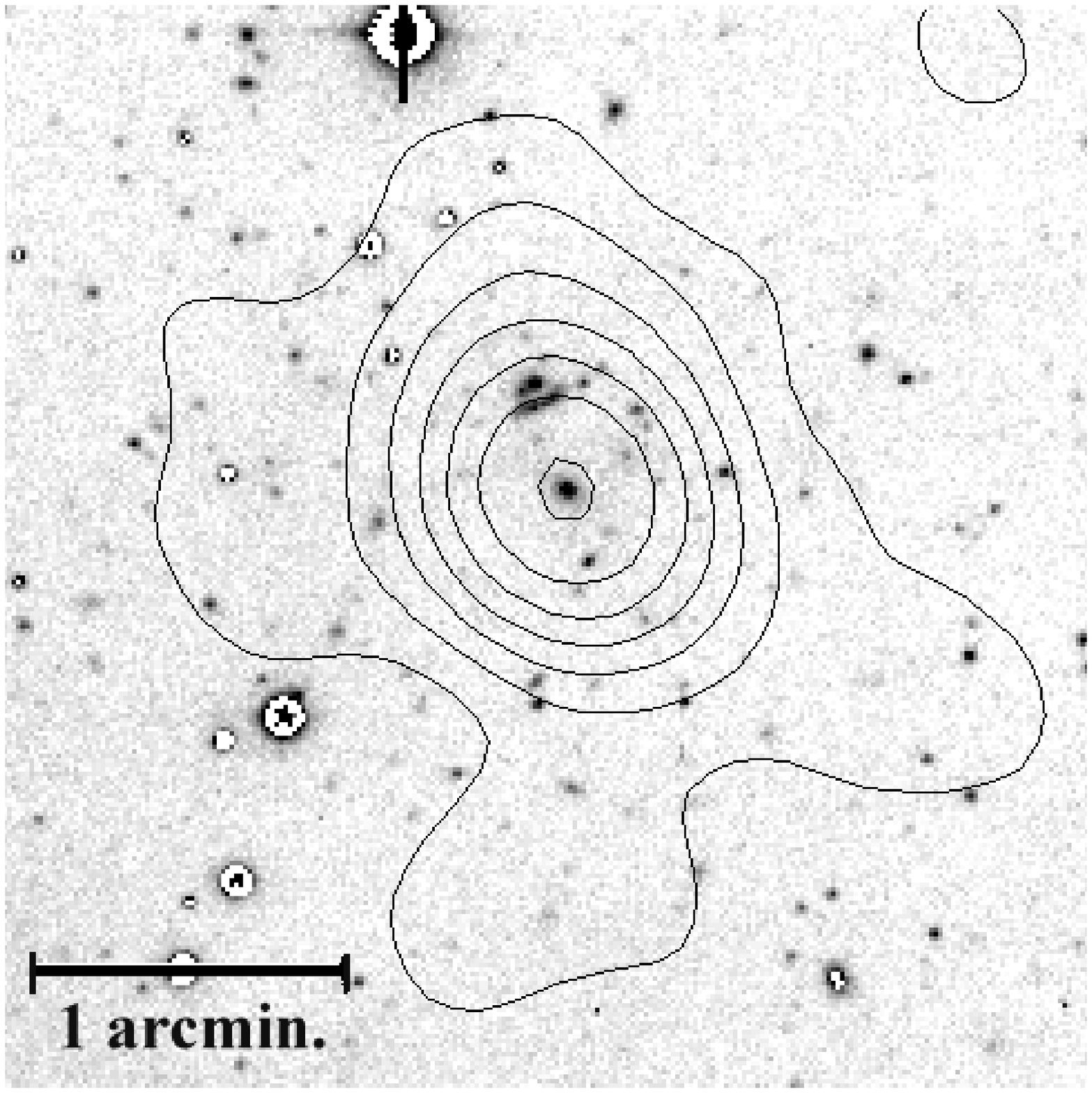  hoffset=-10 voffset=220 hscale=0.4 vscale=0.4}
\caption{The V+R image of CL2236-04 at $z=0.55$ 
taken by ESO 3.6m \cite{Melnick93} 
superposed on the X-ray image obtained by {\sl ROSAT}/HRI.\cite{HAT98}
The straight arc at $z=1.16$, as bright as the bright cluster member 
galaxies, can be observed at the north of the cluster core. 
A bright cluster member galaxy exists close to the arc.
}
\label{fig:cl2236}
\end{wrapfigure}

Hattori et al.\cite{HAT98} showed that in the distant cooling flow
cluster CL2236-04, lensing and X-ray masses agree if the lensing
mass model includes a second clump of mass centered on the second
brightest member (Fig. \ref{fig:cl2236}).  This successful lens model
was first proposed by Kneib et al.\cite{KN94} prior to any X-ray
observations.

Similarly, Hashimotodani et al.\cite{Hashi99b} have shown that there
is no discrepancy if a secondary potential due to the bright cluster
galaxy near the arcs of MS0302.7+1658 and MS1621.5+2640 is included in
the lensing model.  If the contributions of the secondary clump are
not taken into account, the X-ray mass of the clusters are a factor of
2--4 less than the required mass to explain the arc positions in these
two clusters.  Therefore, the introduction of the secondary galaxy
potential is essential.

In the case of the above mentioned clusters, the nature of the galaxy potentials are
constrained consistently with their optical properties.  The strong
lensing in CL2236-04 and MS0302.7+1658 is an example of marginal lensing
(lips in CL2236-04 and beak-to-beak in MS0302.7+1658).

\subsection{A closer look at individual cases}

It seems that the discrepancy could be explained by different
arguments.  To get a better feeling of the current limitation on the
data, we now describe in detail some peculiar cases.

\subsubsection{MS1621.5+2640}

The situation in MS1621.5+2640 is typically a barely constrained cluster
lens.  The observed arc A1 configuration can be reproduced by
qualitatively different mass distribution models.\cite{Hashi99b}  The
first model assumes that the ICM is isothermal and in hydrostatic
equilibrium.  The second model assumes that the dark matter
distribution has a small core radius of $\sim50$ kpc but is much
smaller than the X-ray core radius of MS1621 ($r^X_c=256$ kpc).  The
last possibility was considered to take into account a possibility similar
to A2218.  As pointed out by Moris et al.,\cite{Moris98}
the large core radius and irregular morphology of MS1621.5+2640 may be
a signature of a disturbed distribution of the ICM due to merger.

Arc A1 can be explained by both models if a secondary galaxy 
contribution is taken into account.  However, we cannot distinguish
between models at this moment --- mainly due to a lack of lensing
constraints.  Deeper lensing/X-ray observations will be important to
understand the discrepancy.

\subsubsection{RXJ1347.5-1145}

The origin and history of the discrepancy for RXJ1347.5-1145 is
somewhat different from others.  Schindler et al.\cite{Schind95} found
two arcs in this field.  Schindler et al.\cite{Schind97} made a 
lensing mass estimation assuming that they are multiple images,
and found that the lensing mass is a factor of three larger than that 
expected from X-ray observations. 
However, it
is now clear that they are single images of different background
galaxies, according to their color difference.\cite{FT97}  Therefore,
the mass estimate made by assuming that the arc position is the same
as the Einstein ring radius is not valid (and this is the case for
other clusters).  Fischer and Tyson\cite{FT97} measured a weak lensing
shear field and found that the mass obtained by a weak lensing mass
reconstruction is a factor of 2 larger than that expected from 
X-ray observations.\cite{Schind97} However, as they do not have any
multiple images, it is very difficult to believe an absolute mass
estimate from weak shear estimates, as this will depend on the assumed
redshift distribution of the background galaxies as well as the
correction factor applied to compensate for the seeing effect.  For
example, if the mean source redshift of the shear field measured by
Fischer and Tyson\cite{FT97} through RXJ1347.5-1145 is as large as 1
(and not $\sim 0.65$, as they assumed) the required weak-shear mass is
reduced by a factor of 2, and both the lensing and X-ray mass
estimation provide consistent results.  A photometric estimate of the
source redshift may in the near future address such problems in more
detail, especially with the new 8--10m class telescope, which offers deep
high quality images in relatively short exposures.

\subsubsection{A370} 

The cluster A370 is one of the best modeled lensing
clusters.\cite{KMFM93,Bez98}  Ota, Mitsuda and Fukazawa\cite{Ota98}
found a factor of 2 discrepancy in the cluster central mass estimation
between X-ray and strong lensing.  However, the assumed cluster center
includes very large error because of the poor {\sl ASCA} position
determination accuracy ($\sim 1'$).  Hashimotodani\cite{Hashi99}
analyzed {\sl ROSAT}/HRI data and obtained interesting results (Fig. 
\ref{fig:a370}). The X-ray emission is peaked at the
center of the southern cD galaxy (see Fig. \ref{fig:a370}).  However,
there is no X-ray peak associated with the northern cD, although the
velocity dispersion and the core radius of the northern clump are
similar to those of the southern clump in the lens model.\cite{Bez98}  The
existence of the northern mass clump has also been confirmed by the weak
lensing mass reconstruction using the {\sl HST}/WFPC2 image,\cite{ume99}
been therefore making its existence certain.  
Hashimotodani revised the
results of the {\sl ASCA} temperature. 
It can be safely
assumed that the temperature measured by {\sl ASCA} is the temperature of the gas
associated with the southern clump only.  
Checking the left- and right-hand
sides of Eq. (\ref{eq:sig0}), they derived $\beta$ and the
temperature, and found them to be consistent with the velocity
dispersion of the lens model.\cite{Bez98} 
The X-ray core radius of
the southern clump is also consistent with the lens model.  Therefore,
for the southern clump, X-ray results are consistent with the mass
distribution of the southern clump predicted by lens modelling.
 
Thus, the problem now, is to explain why no X-ray emission is detected
in the northern clump. Is this due to merger?
May the gas around the northern clump and the southern clump just have merged?
But then why did the southern clump not suffer from the merging?
It is likely that A370 is the superposition
of two clusters with slightly different redshifts, as suggested by their
velocity histogramme. However, this is not a fully satisfactory
explanation, as very few X-rays are detected in the Northern clump.

Spatially resolved X-ray spectroscopy with {\sl AXAF} and {\sl XMM} 
observations and a deep wide field spectroscopy survey similar to
the Cl0024+1654 survey will provide exciting new insight for understanding 
cluster formation.

\subsubsection{Is there really a discrepancy?}

It is now clear that comparison between strong lensing and the X-ray mass
must be done carefully from cluster to cluster, based on the detailed
X-ray image and galaxy distributions.  However, we must note that,
except for a few well studied clusters as introduced in this section, the
lensing studies are sometimes not done very accurately --- especially
when they estimate the mass from the Einstein radius.  Since the
measurement of the mass of the cluster with strong lensing events is
based on model fitting, if the model introduced  is not
motivated physically, it may lead to a large systematical error in mass
estimation.

Further work on the mass discrepancy between X-ray and lensing studies should
consider seriously the complexity of the mass distribution
(This is in particular true for the lensing estimate, but it also applies to
X-ray analysis.) before reaching a hasty conclusion on an
unlikely mass discrepancy.

\subsection{Implications}

In this section, we summarize what we have learned from 
strong lensing studies of the cluster mass. 

\subsubsection{Radial arcs and the very central mass profile}

As explained in \S 2, the radial arcs impose a strong constraint
on the degree of central mass concentration in the cluster.  Clusters
with a radial arc are listed in Table I.  A good example of
a radial arc is shown in Fig. \ref{fig:ms2137}.  Radial arcs can only
be seen on the HST images for A370$/$MS0440$/$AC118, as they are very
thin!  The existence of the radial arcs in these systems favors a
central surface mass density distribution flatter than
$\propto\theta^{-1}$ (a mass density distribution flatter than $\propto
r^{-2}$). The thinness of the radial arcs would indicate a
relatively cuspy mass profile; that is $0.5<\delta<1$, where
$\Sigma\propto\theta^{-\delta}$.  The expected slope of the surface
mass density is steeper than that of NFW.  This may suggest that the
radial mass density profile of the cluster central region is steeper
than $r^{-1}$, as indicated by the higher resolution simulation using
GRAPE,\cite{Fukushige97} which has a higher resolution than NFW's simulation.  
However, the central
giant elliptical certainly has influence on the width of radial arcs.
Therefore, the detailed quantification of the tangential deformation
rate of the radial arcs are very important.  Resolving the width,
measuring the redshift and the velocity dispersion of the source
galaxies of radial arcs, and measuring the velocity dispersion profile
of the central elliptical galaxy will be key observations of the
10m class telescopes, such as Keck, SUBARU, VLT, Gemini and in the
moderate future NGST.
\begin{figure}
\vspace{12cm}  
\special{epsfile=ms2137.ps  hoffset=20 voffset=390 hscale=0.7 vscale=0.7}
\caption{The central part of cluster MS2137-23 seen with {\sl HST}/WFPC2/F702W.\cite{Ham97} 
A radial arc is clearly seen. }
\label{fig:ms2137}
\end{figure}

\begin{table}
 \caption{
List of detected radial arcs.  The first column contains the names
of the clusters and radial arcs.  The paper in which  the
radial arc was first reported is cited. Since the radial arcs are elongated along the
radial direction by about a few arcsec, 
the distance to nearby cD is measured from the 
approximate arc center and is shown in the second column.
The third column gives the redshift of the clusters. 
None have a spectroscopic redshift yet. The last column
is the model prediction of the arc redshift.}
  \label{table:1}
  \begin{center}
   \begin{tabular}{cccc} \hline \hline
   cluster (arc name) & dist to cD & $z_{\rm d}$ & $z_{\rm pred}$\\ \hline
   MS2137-23 (A1$)$\cite{Fort92} & $\sim 5^{\prime\prime}$ & 0.313 & - \\
   A370       (R)\cite{Smail96} & $\sim8.''4$ & 0.37 & $1.3\pm0.2$\cite{Smail96} 
\\
   MS0440+02 (A16)\cite{Gioia98}& $\sim6''$ &0.19& - \\
   MS0440+02 (A17)\cite{Gioia98}& $\sim6''$ &0.19& - \\
   AC114    (A4-A5)\cite{NAT98} &  $\sim2.''7$&0.31& $1.76\pm0.15$\cite{NAT98} \\
   AC118  \cite{Kneib99p}& $\sim5''$ &0.33& - \\
   AC118  \cite{Kneib99p}& $\sim5''$ &0.33& - \\
   CL0024-1658\cite{Kneib99p}& $\sim10''$ &0.39& - \\ \hline 
   \end{tabular}
  \end{center}
\end{table}

\subsubsection{The cooling flow interpretation}

Important implications for the cooling flow model can be extracted from
the studies of A2390, CL2236-04 and MS0302.7+1658.  These clusters
have a high central electron density and a short radiative cooling time
and therefore are expected to have cooling flow.  Cluster mass
estimates from X-ray observations using {\sl ASCA} data with a single
phase model fitting\cite{HAT98} are consistent with the strong lensing
estimate.  If the effect of the cooling flow is significant, the
temperature of the hottest phase gas in these clusters could be a
factor of 2--3 higher than the single phase model fitting results.
However, these two clusters are marginal lenses, and the lens models
are very sensitive to the total mass, and therefore to the cluster
temperature.  Increasing the temperature by more than factor of 2 is
incompatible with the existence of a straight arc in these clusters.
Therefore, even if cooling flow exists in these clusters, the
deviation of the hottest phase temperature from the {\sl ASCA}
measured temperature must be negligible.

\subsubsection{Existence of a universal mass profile}

The two models that can explain the observed arc configuration of
MS1621.5 +2640 have very important but very different cosmological
implications.  If the hydrostatic model is correct, MS1621.5+2640 has
a very large core radius and a low central electron density, 
respectively a factor of 5 larger and 10 less than those of
MS0302.7+1658.  MS1621.5+2640 is known to be a non-cooling flow cluster, since its
central cooling time is longer than the Hubble time, although the
temperature of MS1621.5+2640, $\sim 6.5$keV, is not much different
from that of MS0302.7+1658, $\sim 5$keV.  This indicates that there
are large variations in the core radius and the central density of the cluster mass
distribution even if the temperature of the clusters have similar values.
This fact has been noticed by Fujita and Takahara,\cite{Fujita99a} who
examined the fundamental plane of nearby clusters.  They concluded
that if the ratio of the dark matter density to the
gas density at the cluster center is constant in all clusters,
the cluster mass distribution cannot be characterized only by its
virial mass and redshift; the formation epoch of the cluster and its
evolutional history may also be important parameters.  Since
Hashimotodani et al.\cite{Hashi99b} are using lens modelling to
estimate the central dark matter density, their results are free of the
assumption on the ratio between the dark matter and gas densities.
Therefore, the standard Press-Schechter theory\cite{PS74} should
be used with caution, as it assumes that the cluster is just
virialized at the observed redshift when the theoretical prediction of the
cluster temperature and luminosity functions are made.  Modifications
of the standard Press-Schechter theory, which take into account the
difference between the formation and observed redshift, are thus
crucial for comparing the theoretical predictions with the observed
cluster temperature and luminosity
functions.\cite{LC93}\tocite{Fujita99b}  Nevertheless, if the compact
core model is correct, it indicates that the central electron density
is not a good representative of the central dark matter density, and
the possibility that the clusters have a universal mass density profile, 
characterized only by its redshift and virial
temperature, may then be close to reality.

It is also worth to note that Bartelmann and Steinmetz\cite{Bart96} claim
that the $\beta$-model does not provide a good description of the X-ray surface
brightness profile of clusters as determined from their
$N$-body-hydrodynamics simulations.  They argue that the $\beta$-values
in the fit depend critically on the background level and that this
introduces a significant bias in the mass determination built on
$\beta$-model fits.  Actually for most of the lensing clusters, the
central X-ray surface brightness is close to the background level of
{\sl ROSAT}/HRI, and the $\beta$ values obtained, $\beta\sim
0.45-0.55$, are systematically somewhat lower than those for nearby
clusters, $\beta\sim 0.6-0.7$.  However, we have seen that the cluster
mass distribution models constructed with these relatively small $\beta$
values are consistent with the cluster mass distribution modeled by
the strong lensing observations.  Although this result may interpret
that the obtained small value of $\beta$ for the distant lensing
clusters represents the real mass distribution well, further X-ray
observations with much higher sensitivities and with similar or better
spatial resolution than {\sl ROSAT}/HRI, e.g. {\sl AXAF}, {\sl XMM},
are required to clarify the radial profile of clusters in the
outskirts.

\subsubsection{Extrapolation of general relativity up to the cluster scale}

We can now draw important conclusions concerning whether cluster lensing provides
empirical support for the extrapolation of general relativity up to the
cluster scale.  If this extrapolation fails on the cluster scale,
then the discrepancy between lensing and X-ray mass estimates should be
systematic.  However, as we have seen, excellent agreement
between these results has already been reported in about half a dozen
clusters.  Clusters still exhibiting a discrepancy are likely to be
understood in an astronomical context when better quality data are in
hand to complete a detailed analysis.  Therefore, we conclude that
the extrapolation of general relativity up to the cluster scale is
empirically supported by cluster lensing.

\section{The missing lens problem and very distant clusters}

Looking for and studying very distant galaxy clusters, clusters with
$z>1$, are very important because: i) the number density of rich
clusters at large redshifts is a very sensitive measure of the mean
density of the Universe; ii) high redshift clusters provide
important information to understand the formation and evolution of
galaxies (e.g. the evolution of the metallicity of the hot intra-cluster
medium, morphological evolution, and star formation history in a dense
environment).

However, this is a difficult enterprise, as there are yet only a few
confirmed very distant galaxy clusters.  The first necessary step is
to identify very distant clusters by conducting very deep
observations.  A random search of the sky is not practical since one expect
to detect only a very small number of very distant clusters.
A possible route to identify high redshift clusters is to search in
a dark lenses field, as explained below.

The expression {\it `missing lens problem'} can be found in the unique textbook
on gravitational lensing written by Schneider, Ehlers and
Falco.\cite{SEF92}  A number of quasar pairs are candidates to be
lensed systems. In a high fraction of these systems, either no deflector or
only a part of a deflector is clearly identified.  The difference
between this problem and the usual missing mass problem is that the
mass-to-light ratio required to explain the pair of quasars by
a gravitational effect is extremely large compared to that of galaxies.
A yet unidentified part of lens objects (distinct from a galaxy
and$/$or galaxy clusters), is called a `dark lens'.  The dark lens
search is an attempt to identify  dark lens objects as already
known objects.\cite{WC96}  If all ultra-deep multi-wavelength searches
fail to find any evidence of galaxies and/or clusters as the lenses, it
could lead us to conclude the existence of a yet unknown new type of
object which contains barely any luminous matter.  Table 2 of
Hattori\cite{HAT97iau} summarizes such lens candidates.  A large
fraction of the lens candidates exhibit the missing lens problem.
We now discuss two of them, for which mysterious rich clusters of
galaxies may play an important role.

\subsection{The lens system MG2016+112}

\begin{figure}
\vspace{18cm}  
\special{epsfile=axj2019.ps  hoffset=-30 voffset=460 hscale=0.8 vscale=0.8}
\caption{Unveiling the nature of the dark cluster AXJ2019+112.  
The I-band image of AXJ2019+112 field taken with {\sl NOT} and overlaid
X-ray contours taken by {\sl ROSAT}/HRI.\cite{HAT97} 
The spectroscopically measured redshifts of galaxies are attached and 
possible member galaxies of the cluster AXJ2019+112 are marked by ``the thick 
circle."  Although a dozen member galaxies were confirmed by deep Keck imaging
studies and photometric studies,\cite{Benitez98}
there is a debate on the interpretation of the {\sl ROSAT}/HRI results. 
}
\label{fig:AXJ2019}
\end{figure}

Hattori et al.\cite{HAT97} have discovered a cluster of galaxies at
$z=1$, named AXJ2019 +1127, using X-ray observations in the direction of
MG2016+112, a confirmed multiple QSO at $z=3.27.$\cite{Lawrence}  When
discovered, this cluster was claimed to be a dark cluster because of
the lack of an optical counterpart,\cite{DSchneid85,Lang91} although the
existence of a massive cluster has been predicted by theoretical
modelling.\cite{Narasimha87}  Recent deep spectroscopic and imaging
studies in optical and IR bands unveiled the nature of the dark
cluster.  Kneib et al.\cite{Kneib97} spectroscopically found 6
possible member galaxies in this field (Fig. \ref{fig:AXJ2019}).  Using
deep $V,\;I$ Keck images, and wide-field $K_s$ NTT images, Ben\'\i tez
et al.\cite{Benitez98} found a tight red sequence of galaxies which
has a slope in good agreement with the model predictions for $z\sim1$
clusters.\cite{Kodama98}  Based on the identified number of member
galaxies\cite{Benitez98} and the mass estimated from the X-ray
temperature,\cite{HAT97} the mass-to-light ratio within an aperture of
$0.8h_{50}^{-1}$Mpc was found to be $M/L_V\sim 110h_{50}$, which is consistent
with the mass-to-light ratio found in nearby clusters.\cite{Bahcall95}
Therefore, it is very likely that a massive cluster of galaxies exists
at $z=1$ in the line-of-sight of MG2016+112.  AXJ2019+1127 is one of
the most massive clusters known at $z\sim 1$. Furthermore, it is 
one of the rare high-redshift clusters for which strong lensing events
are clearly detected. Thus, it provides a unique opportunity to study
the mass distribution in a very high redshift cluster.  Ben\'\i tez
et al.\cite{Benitez98} successfully modeled (with a cluster surface
mass density distribution similar to the NFW model) the multiple
images of QSO and the extended arc-like image detected in the {\sl
  HST}/NICMOS image.  The cluster mass derived from the lens modelling
is also consistent with X-ray results.  

It is worth noting that the
Einstein ring radius of AXJ2019+112 is about $2''$\cite{DSchneid85} if
the cluster centre coincides with the giant elliptical located at the
middle of the triple QSO images, and it is very small compared to
clusters at intermediate redshift ($z\sim 0.2-0.5$), which are as hot
as AXJ2019+112.  Therefore, the central surface mass density of
AXJ2019+112 is shallower than those cluster lenses with similar
temperatures at intermediate redshift.  This may be reflecting the situation in which 
1) there is a large variation in the central surface mass density from
cluster to cluster independent of their redshift, or  2) there is
a strong evolution in the central surface mass density.  However, it
should be emphasized that the interpretation of the {\sl ROSAT}/HRI
results of AXJ2019+112 leads to different
results.\cite{HAT97,Benitez98}  Although Ben\'\i tez et
al.\cite{Benitez98} claimed that there are three X-ray sources in this
field, since there are three patches in the X-ray image, it is dangerous to
take these as real sources.  For example, the early {\sl ROSAT}/HRI
image of A370\cite{Fort94} exhibited two X-ray peaks coinciding with the
centre of the two giant ellipticals.  However, as shown in
Fig. \ref{fig:a370}, the most up-to-date X-ray image of A370 (with an
exposure time twenty times longer than the previous one) does 
show only one prominent peak.  Clearly, photon statistical errors must
be considered seriously before reaching hasty, unjustified conclusions.
Therefore, determination of the accurate cluster central position,
morphological studies by detecting more than an order of magnitude larger
number of X-ray photons, and the accurate measurement of the X-ray
temperature of AXJ2019+1127 will be key observations to conduct with
future X-ray missions.

\subsection{The double quasar Q2345+007}

The quasar Q2345+007 has been one of the most mysterious lens
candidates of double quasars since its discovery.\cite{Weedman82}  In spite
of deep and wide field optical searches, the main lens
 has not yet been identified.\cite{Tyson86}  A very faint galaxy
was found at the edge of the secondary image, B image, after  subtraction
of the point spread function of the image.\cite{Fischer94}  Since a C
IV doublet absorption line ($z=1.483)$ \cite{Foltz84,Steidel91} was
found in the B image, the redshift of the faint galaxy is supposed to
be 1.483. However, the expected mass of the galaxy is too small to
explain the wide separation of the two images, unless it has an
extremely large mass-to-light ratio.  A large cluster at $z=1.49$ was
claimed, since both of the two images have metal absorption lines with
redshift $z=1.491$.  Bonnet et al. \cite{Bonnet93} reported the
detection of a possible lensing cluster from a weak lensing shear field
with a strength of $\gamma \sim 0.15$, confirmed by van Waerbeke et
al.\cite{vanWaerbeke} and arclet candidates.  After this detection, a
galaxy cluster candidate was found as an enhancement in the number
density of faint galaxies at the right location in the sky predicted
by the shear field.\cite{Fischer94,Mellier94}  Pell\'o et
al.\cite{Pello96} made photometric redshift estimations for the
galaxies in the cluster candidate field and found an excess of
galaxies at $z\sim 0.75\pm 0.1$.  Assuming the cluster has this
redshift, the velocity dispersion required to produce the observed
shear pattern should be $790^{+115}_{-170}{\rm km/s}$,\cite{Pello96}
which corresponds to a hot gas temperature of $k_{\rm
  B}T\sim4.1^{+1.3}_{-1.6}(\beta/1.0)^{-1}$keV.  The typical value of
$\beta$ ($\sim 0.66$) yields $k_{\rm B}T\sim 6.2$keV, which is as hot
as CL2236-04 at $z=0.552$.  Therefore, it is expected that the cluster
speculated from the lensing and photometric studies is a bright X-ray
source.  However, non detection of X-ray emissions from the cluster by
{\sl ASCA} and {\sl ROSAT}/HRI observations was
reported.\cite{HAT97iau} This deepened the mystery
further.

\subsection{Implications}

Other than the original interest in the missing lens problem (that is that
it indicates the existence of a yet unknown new type of dark objects in
the universe), it is likely that the dark lens search is an
efficient way to detect a very distant massive cluster and provides
a unique opportunity to study mass distributions in very distant
clusters.  Therefore, searches for massive condensations in the field
of  dark lens objects by the detection of the lensing signatures
with deep optical imaging and by  X-ray detection is
a possible direction for observational cosmology in the near future.

\section{Clusters of galaxies as natural gravitational telescopes}

\subsection{Looking at distant galaxies through cluster-lenses}

Cluster lenses magnify and distort galaxies behind them. For
an efficient cluster (a massive cluster with intermediate redshift $z\sim
0.2$--$0.4$) the magnification factor for the faint galaxy population is
typically $\sim 2$ in a few square arc-minutes. This gain would
correspond to a factor of $\sim 1.5$ in the diameter of a telescope or an
increase by a factor of $\sim 4$ in exposure time. Clearly, looking
through cluster lenses can be of some reward when studying a faint
(and thus distant) galaxy population, as it allows us to observe
intrinsically fainter objects that would not otherwise be observable
when looking at blank fields.  Cluster lenses do magnify but
also distort the shape of distant galaxies, and the farther the source,
the stronger the distortion. Hence the shape of a lensed galaxy
(and whether it is multiple or not) is generally a good distance
indicator.  These properties have effectively been used in detail
in recent years to study the faint and distant galaxy population. 
Thus cluster lenses have been nicknamed ``natural telescope''.

Cluster lenses can be used as telescopes in two different ways, 
i) as  magnifying glasses for non-resolved sources. Here, only the
magnification property is used, to constrain galaxy counts
to fainter limits, and to study photometric properties of fainter
galaxies.   ii)Cluster lenses can also be used 
as real telescopes for resolved sources. Here,
both  magnification and distortion properties are used to  1)
constrain galaxy morphology to fainter limits, and 2) constrain their
distance through their observed shape.

The high quality imaging obtained with the {\sl HST}/WFPC2 camera
provides the {\it key} observations that make possible the use of
cluster lenses as natural telescopes. Indeed, thanks to its
high-resolution imaging capabilities it allows us to better identify
morphologically multiple images and to accurately compute the shapes of
arclets. These critical parameters are the necessary ingredients for
the construction of a robust mass distribution model of cluster lenses.
Furthermore, if at least one multiple image has a spectroscopic
redshift, then the mass distribution model is absolutely calibrated,
making a cluster lens an effective cosmological telescope.

\subsection{Resolved lensed galaxies --- Morphological and physical properties}

\begin{wrapfigure}{r}{6.6cm}
\vspace{7cm}  
\special{epsfile=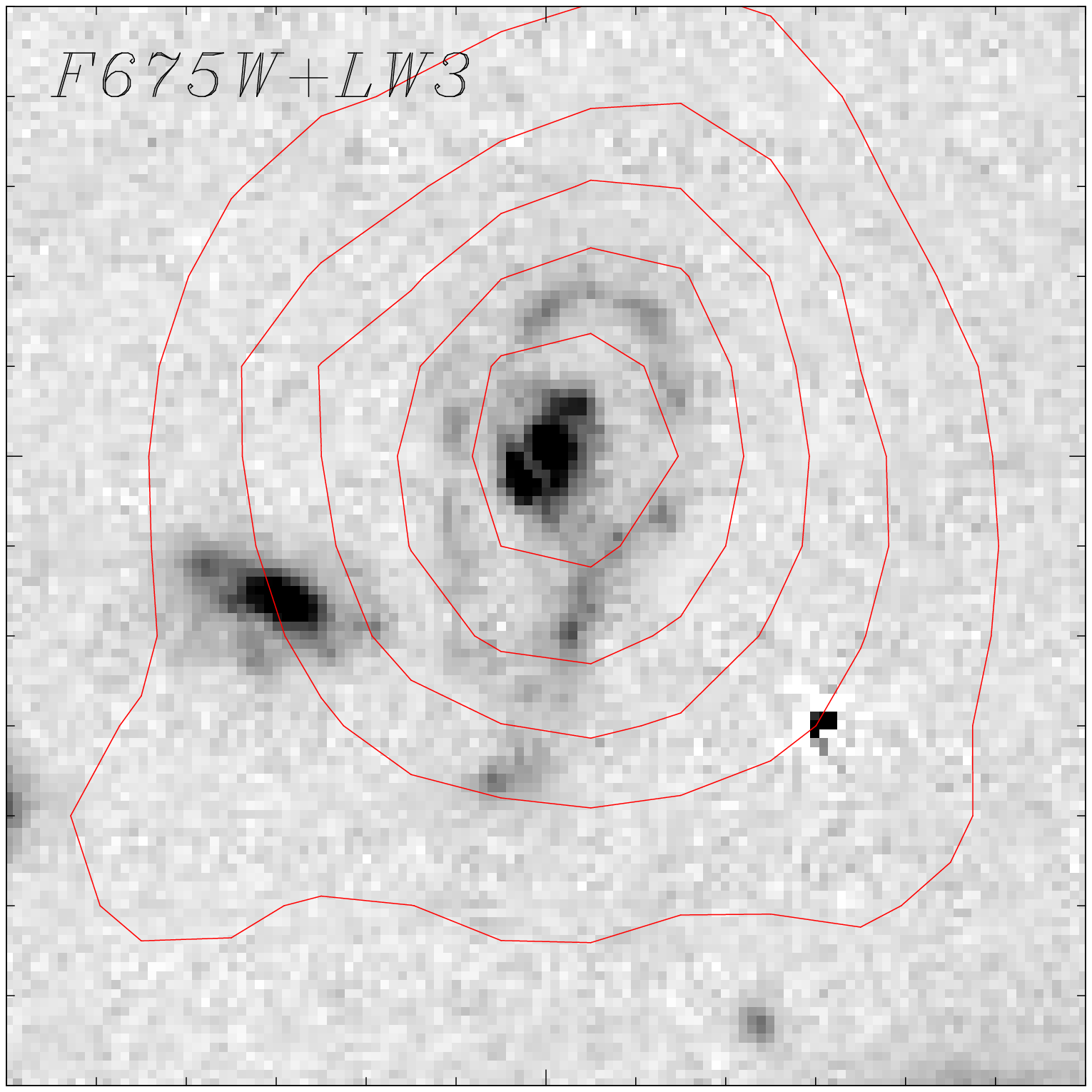  hoffset=8 voffset=210 hscale=0.4 vscale=0.4}
\caption{The image of the distant ring galaxy, LRG J0239-0134 
at $z=1.062\pm 0.001$ magnified by cluster lens A370.\cite{Soucail99}
{\sl ISO}/LW3 contours overplotted on {\sl HST}/WFPC2 F675W image (with a $12''$ size). 
This object is appeared closed to the galaxy \#32 but this galaxy has 
been subtracted for clarity (a black dot position at the right of the ring galaxy). 
}
\label{fig:LRGJ0239}
\end{wrapfigure}

Morphological properties of distant lensed sources were first
addressed by Smail et al.,\cite{Smail96} who estimated the intrinsic
linear sizes of the galaxies and showed them to be compatible with a
significant size evolution with redshift.  Smail et al. \cite{Smail96}
studied the {\sl HST} images of four distant massive clusters and
found that the half light radius of the arcs along the radial
direction of the cluster potential is a factor of 1.5--2 times smaller
than the equivalent population in the local field.  If the gross
properties of the central regions of the cluster surface mass density
at the position of the tangential arcs can be characterized by
$\Sigma\sim \theta^{-1}$, the radial stretching rate is unity (see
\S 2.1), and the above mentioned results can be taken as the
intrinsic nature of the arcs.  Since they showed that the $\Sigma\sim
\theta^{-1}$ is a good representative of the central mass distribution
of their sample clusters, they suggested that the dominant population
of star-forming galaxies at $z\sim 1$ is a factor of 1.5--2 times
smaller in size than the equivalent population in the local field.  If
the cluster surface mass density profile obtained at the radial arc
position can be extrapolated to the tangential arc position, the
surface mass density is shallower than $\Sigma\propto\theta^{-1}$, and
the radial width of the tangential arc is larger than the intrinsic
source size.  This strengthens the conclusion reached by Smail et
al.\cite{Smail96}  Recently, a number of high-$z$ arcs and arclets
have been confirmed spectroscopically.\cite{Mellier98,Ebbels96}
\tocite{Frye98}
A detailed analysis of their intrinsic size and morphology is then
possible using lens modelling. 
More recently, several detailed analyses of the morphology of 
high-$z$ arcs were reported using lens modelling.
For example, Soucail et al.\cite{Soucail99a} found a ring galaxy LRG
J0239-0134 at $z=1$ through the cluster lens A370
(Fig. \ref{fig:LRGJ0239}).  The morphological nature of this object is
more complex than that of nearby normal galaxies and is very similar to the
Cartwheel galaxy,\cite{Struck96} which is thought to be an interacting
galaxy.  Most of them appear to be knotty with a more complex
morphology than their local counterparts,\cite{Colley96}
\tocite{Pello98} although there may be some biases due
to observations in the UV rest frame.

\subsection{Resolved lensed galaxies --- Distances of  faint populations}

The probability distribution of the redshift of an arclet for a given mass
distribution of the cluster and the shape of the arclet depends only
on the intrinsic ellipticity distribution of faint galaxies:
\cite{Kneib96} the larger the redshift, the larger (in general) the
deformation induced by the cluster. From a secure mass distribution
defined by a few sets of multiple-images, we can then estimate a
likely value of $z$ of arclets behind well-constrained clusters.  This is most
interesting, as this method is purely geometrical and therefore does
not suffer from the spectroscopic bias that provides redshifts only
for these faint galaxies with (strong) optical emission lines.  To
evaluate the accuracy of this technique on a cluster-lens {\bf A2218},
\cite{Kneib96} Ebbels et al.\cite{Ebbels98} have successfully
measured with the 4.2m {\sl WHT} the redshift of 19 arclets, providing
a first confirmation of the lensing method.  Their results shows that
the mean redshift of the faint field population at $R\sim 25.5\;
(B\sim 26$--$27)$ is low, $\langle z\rangle=0.8$--$1$.  Similar work is now in progress
for other cluster lenses, such as {\bf A2390}\cite{Kneib99} and {\bf
  AC114}.

\subsection{Non-resolved lensed galaxies in the Submm and MIR bands}

\begin{figure}
\vspace{13.5cm} 
\special{epsfile=scubant.ps  hoffset=30 voffset=390 hscale=0.8 vscale=0.8}
\caption{Cumulative sub-mm number counts through cluster lenses and in the field.\cite{Blain98}
The results obtained through lenses are indicated by solid squares. 
The results obtained by earlier works from Barger et al. (B98),\cite{Barger98} 
Ealse et al. (E98),\cite{Ealse98} 
Holland et al. (H98),\cite{Holland99} Hughes et al. (Hu98),\cite{Hughes98} 
and Smail et al. (SIB)\cite{SIB97} are shown by 
solid circles.  
The number counts through cluster lenses reaches down to a very faint flux density 
limit of 0.5mJy at 850$\mu$m, several times fainter 
than the confusion limit of the {\sl JCMT} in blank fields.\cite{Hughes98}
}
\label{fig:scuba}
\end{figure}

A fruitful avenue that has developed in the last two years is the
study of distant galaxies in until recently unexplored windows of the
electromagnetic spectrum: the Mid-IR, the Sub-millimetre and
Millimetre bands. Indeed, the ISOCAM camera on the {\sl ISO} satellite and
the SCUBA instrument on {\sl JCMT}, thanks to their larger survey efficiency
(compared to previous instruments), allow for studies of distant galaxies.
Boosting their detecting efficiency with lensing magnification has
shed new light in the nature of faint galaxies detected in these
two wavebands (Fig. \ref{fig:scuba}).

Distant, $z>3$, objects were supposed to be strong sub-mm sources
because we do expect to detect redshifted dust-emission from these
distant galaxies.\cite{Blain97}  A first tentative application in
sub-mm has been recently presented by Smail, Ivison and
Blain.\cite{SIB97}  It is now extended to cover seven
clusters.\cite{Smail98a}  They found in total 16 3$\sigma$ sources and
10 4$\sigma$ sources at $850\mu$m through all seven clusters.  Their
sub-mm galaxy counts greatly exceed those expected in a non-evolving
model based on the local {\it IRAS} $60\mu$m luminosity function
\cite{Saunders90} or in a model that can explain the optically
selected Lyman-dropout\cite{Madau96} and Lyman-emission
samples\cite{Hu98}(Fig. \ref{fig:scuba}).  Smail et al.\cite{Smail98b} 
have attempted to find
optical counterparts of these sub-mm sources using {\sl HST} images.
Counterparts were identified for 14 of the 16 3$\sigma$ sources.  One
of them is thought to be the distant ring galaxy LRG J0239-0134 shown
in Fig. \ref{fig:LRGJ0239}.  The morphologies of the optical
counterparts fall into disturbed$/$interacting objects and compact
objects.  The disturbed and interacting galaxies constitute the
largest class, which suggests that interactions play an
important role in triggering star formation and nuclear activity in
the distant universe.

A deep survey through cluster lenses in the mid-IR wave band ($7\mu$m
and $15\mu$m) using {\sl ISO}/ISOCAM also has been
performed.\cite{Metcalfe99}  Currently, a total area of about
53 arcmin$^2$ has been covered in maps of three clusters.  The source
counts, corrected for cluster contamination and lensing distortion
effects, exhibit an excess by a factor of 10 with respect to the
prediction of a non-evolution model.\cite{Altieri98}  The results
suggest that abundant star formation occurs in very dusty environments
at $z\sim 1$.  As shown in Fig. \ref{fig:LRGJ0239}, the distant ring
galaxy LRG J0239-0134 is a bright mid-IR source.\cite{Soucail99a}  The
fact that the [OII] line is more extended than the underlying stellar
continuum indicates on going star formation in the ring.  The
detection of the [NeV] emission line indicates the existence of an
active nucleus.  These results indicate that a galaxy-galaxy interaction
triggered the formation of the star-forming ring and nuclear activity
in LRG J0239-0134.

These results indicate that interactions play an important role
in triggering star formation and nuclear activity in the distant
universe, and that care must be taken to take account of the
dust in inferring global star formation history from UV and optical
luminosities of high-$z$ galaxies.

\section{Arc statistics}

As in the case of the statistical study of the multiply imaged
QSOs,\cite{Chiba99} it is natural to expect that the statistical study
of arcs could be a powerful tool to
study\cite{NemDek89,Miralda93a,Miralda93b,GROS94}  i) the cosmological
world models (and particularly to the cosmological density parameter $\Omega_0$ and 
the cosmological constant $\Lambda_0$), ii)
the average cluster mass distribution/profile, and iii) the
nature and evolution of high redshift galaxies.  Aiming to extract
fruitful information regarding these topics, statistical studies of arcs
have been performed by many authors. These studies have been referred to as
``arc statistics".  Clearly, arc statistics depend on a large number of
parameters which still contain large uncertainties, and we therefore
here attempt to clarify its possible application in order to sharpen the focus for its
future use.

Wu and Hammer\cite{WuHam93} provided the definition for a
giant luminous arc (GLA) and performed the first quantitative
comparison between the theoretically predicted number of GLAs and the
observed number of GLAs.  Because no observational results on
systematic arc surveys had been reported, the frequency of finding GLAs in
clusters was extrapolated from the reported number of GLAs at that
time.  They found that the frequency of finding GLAs in clusters is
very sensitive to the degree of the central mass concentration in the
clusters.  They concluded that the observed high frequency of finding
GLAs indicates a high central mass concentration in the clusters.  The
first well-defined GLA survey was conducted by Le F\`evre et
al.\cite{Lefevre94}  They found 6 GLAs among 15 X-ray luminous
clusters. (Although their original sample contained 16 clusters, one of
their samples turned out not to be a cluster.\cite{Mory98})  They
selected X-ray luminous clusters with $z>0.2$ from the EMSS.  The first
quantitative comparison with these survey results was performed by
Hattori, Watanabe and Yamashita.\cite{HWY97}  They found that
the theoretically expected number of GLAs within the framework of the
circularly symmetric cluster potentials, which is consistent with X-ray
observations, is two orders of magnitude less than the observed number.
Molikawa et al.\cite{Mory98} found that the radial profile of Le
F\`evre et al. sample clusters have a large variation from cluster to
cluster, and they are far from the radial profile adopted by Hattori,
Watanabe and Yamashita.\cite{HWY97}  However, Molikawa et
al.\cite{Mory98} reached the same conclusion as Hattori, Watanabe and
Yamashita.\cite{HWY97}  Hamana and Futamase\cite{Hamana97} examined
the GLA statistics by taking into account the evolution of the
luminosity function found by the Canada-France Redshift
Survey.\cite{Lilly95}  They showed that the observed evolution in the
galaxy luminosity function causes the GLA number to increase by at most a factor
of 2--3.  These results suggest that more realistic modelling of
the lens cluster potential, taking into account irregularities due to
substructures and bright member galaxies, is important to account for the
discrepancy.  The importance of the irregularities is also claimed in
the series of works by Bartelmann and his
collaborators\cite{Bart94}\tocite{Bart98} using numerical simulations.
They\cite{Bart95} found that the irregular distribution of dark matter
in clusters can increase the cross section of forming giant arcs by two
orders of magnitude compared with the spherical symmetric mass distribution
model.

As for the next step of the arc statistics, if one considers  the
explosive progress in the refinement of the determination of the mass
distribution in individual clusters by strong and/or weak lensing
observations, constraining the mass
distribution of clusters using GLA statistics may not be an efficient way.  The arc statistics
should be studied with the well-constrained mass distribution
model of the individual cluster by the detailed lensing studies best
with {\sl HST}/WFPC2 data, with the help of X-ray data.  This process
must be done for at least the 15 clusters in Le F\`evre et al. sample,
because this sample is an unbiased sample and the best for a statistical
treatment.  Then the questions to be addressed using the arc statistics
are those concerning the evolutionary nature of the high redshift galaxies and
maybe these involved in constraining the cosmological world model.  This kind of work has
already been initiated by B\`ezecourt.\cite{Bez98b}
B\`ezecourt studied the statistics of an arclet in A370
because the cluster mass distribution of A370 is well-constrained by
lensing.  It was shown that using a realistic cluster potential is
essentially important to recover the observed number of  arclets.

Several meaningful results for probing the high redshift universe using
arc statistics have been reported.  Miralda-Escud\'e\cite{Miralda93b}
showed that the number of GLAs responds sensitively to the intrinsic
size of galaxies, and it decreases drastically when the intrinsic size
of galaxies becomes smaller than the seeing FWHM.  Hattori, Watanabe and
Yamashita\cite{HWY97} showed that the apparent-unlensed source size
distribution of GLAs has a sharp peak at the seeing FWHM.  The seeing
FWHM of the current arc survey of $0.''8$\cite{Lefevre94} is about the size
of the $L_*$ spiral galaxies at $z\sim1$ if there is no evolution in the
source size, and the drop of the number in the larger source size is due
to the exponential decrease of the galaxy luminosity function over $L_*$.
Therefore, it is expected that the source galaxies of GLAs are mostly
the spiral galaxies with $z<1$.  Wu and Hammer\cite{WuHam93} and Hattori,
Watanabe and Yamashita\cite{HWY97} confirmed this.  Hattori, Watanabe and
Yamashita\cite{HWY97} pointed out that the fraction of GLAs which are
spheroidal galaxies nearly in their forming epoch in the observed GLAs
is relatively large 
if the SED of the UV bright elliptical is a good
representative of the majority of elliptical galaxies.  It is
worth  noting that a fraction of the GLAs and multiply imaged galaxies
are made from the high redshift galaxies with
$z>2$.\cite{Yee96,SEIT98,Pello98}  This may suggest either a lacking in our current
understanding of the evolutional features of the high redshift galaxies
or that high redshift elliptical
galaxies in the formation process started to be detected as GLAs.
Quantification of the star formation rate in the GLAs by multi-band 
observations, especially in MIR and sub-mm wave bands mentioned 
in \S 5,  may answer to this question soon.

A more extended arc survey for an unbiased sample containing a much
larger number of clusters is of course important.\cite{Luppino98}  The
cluster sample provided by the {\sl ROSAT} all-sky survey is
undoubtedly a good unbiased sample for the arc survey.  It is also
expected that a new, better cluster sample will be provided by an 
all-sky survey in the 2--10keV band, which is more appropriate for detecting
clusters than the {\sl ROSAT} band, planned to be performed by the X-ray
satellite {\sl ABRIXAS}\cite{ABRIX} soon.

Topics in connection with cosmological constraints are
presented in the next section.

\section{Cosmological constraint by cluster lensing}

\subsection{Constraints from arc statistics}

Bartelmann et al.\cite{Bart98} performed a set of cosmological 
$N$-body simulations, and found that 
the optical depths for the formation of giant arcs is sensible 
for the cosmological world models and is the largest 
for the low $\Omega_0$
cold dark matter model (OCDM).  
They concluded that the OCDM is likely to be the only  model which can match
current observations of the giant arc survey.  
These differences originate from the different epochs of cluster 
formation among the cosmological models.  
They found that  clusters with $z\sim 0.2$--$0.6$, which are the most responsible 
for the formation of arcs, in the OCDM exhibit
the most irregular shape in mass distribution, and as a result they have the 
strongest shear, since the cluster formation   
epoch is just around this redshift range in this model. 
This is very interesting and a practical variation of the idea that uses the degree of 
inhomogeneity in cluster mass distributions to constrain $\Omega_0$ first proposed by 
Richstone, Loeb and Turner.\cite{Rich92}  One of the weakness of this kind of method 
is that the observable, that is the degree of inhomogeneity, to be compared with the model prediction
is somewhat abstract.  
However, Bartelmann et al.\cite{Bart98} made quantitative comparison possible 
by connecting with arc statistics. 
However, the contribution of the bright member galaxies are not taken
into account because of their numerical resolution limit, although 
they clearly have central role in cluster strong lensing, as shown in \S 3.
Further improvement of numerical simulations is required.  

The redshift distribution of the GLAs have been proposed 
to constrain the cosmological constant.\cite{WuMao96}
However, the  redshift distribution of the GLAs is very 
sensitive to the mass distribution model of the cluster. It requires 
a very accurate modelling of the cluster mass distribution  
to constrain 
the cosmological constant.
Further, the expected number of  GLAs in one cluster 
is less than 1. We must perform an arc survey for a large number 
of clusters for which the mass distribution is precisely constrained.  
This will be a difficult task, but it is required to be done.

\subsection{Cosmological constraints using $(\Omega_0,\Lambda_0)$ 
dependence of $D_{ds}/D_{s}$ and $\omega(z)$}

The combination of the angular diameter distance appears in the lens
equation as $D_{\rm ds}/D_{\rm s}$, and its value depends only on the
cosmological constant.\cite{Miralda91}  Using this fact, several
methods to constrain the cosmological constant have been proposed.
The merit of this kind of method is that one cluster is enough to
provide a stringent constraint on the cosmological constant.  However,
the demerit is that the mass distribution of the cluster must be
measured very precisely.  One variation of this method is using the
lensing factor $\omega(z)$ defined as
\begin{equation}
\omega(z)\equiv {D_{\rm ds}(z)/D_{\rm s}(z)\over D_{\rm ds}(\infty)/D_{\rm s}(\infty)}.\label{eq:wz}
\end{equation} 
The convexity of the lensing factor has a $(\Omega_0,\Lambda_0)$
dependence.  Gautret, Fort and Mellier\cite{Gautret98} proposed the
triplet method, in which $(\Omega_0,\Lambda_0)$ is constrained by using
this dependence.  The advantage of the triplet method is that it does
not require the accurate measurement of the cluster mass distribution.
Although the Freedman distance is used as the angular diameter
distance in these methods, it is likely that there is a variation of
the angular diameter distance from line-of-sight to line-of-sight due
to the inhomogeneity of the universe.\cite{Hamana99,TAH99}  Therefore, the
quantitative calibration of the variation of the angular diameter
distance is an important complementary works to extract reliable results from these
methods.  In this section, three methods proposed to constrain
$(\Omega_0,\Lambda_0)$ following the above mentioned principle are summarized.

\subsubsection{Constraints from depletion curve}

Broadhurst\cite{Broad95} first found that the galaxy number count
in the direction of the cluster  exhibits diagnostics which are now called
the ``depletion curve".  Since the detail nature of the depletion curve
depends on the magnification due to the cluster lens, he proposed to use
it for measuring the cluster mass.  Fort, Mellier and
Dantel-Fort\cite{FMDF96} pointed out that the depletion curve can be
used to constrain the cosmological constant if the mass distribution
of the cluster is accurately measured.  The expected galaxy number
density in the direction of the lensing cluster is
\begin{equation}
N(r)=N_0 \mu (r)^{2.5\alpha -1},
\label{eq:27}
\end{equation}
where $\mu$ is the magnification at the position $r$ and $\alpha={\rm
  d\;log}(N)/{\rm d}m$ is the slope of the galaxy number counts.  The
first term of the exponent, $2.5\alpha$, describes the increase of
the galaxy number density, since fainter galaxies become
brighter than the limiting magnitude due to the lensing magnification.
The last term of the exponent, $-1$, describes the decrease of the
galaxy number density, since the unit solid angle area of the unlensed
source plane becomes a factor of $\mu$ larger due to lensing, and the
galaxy number within the unit solid angle of the lensed source plane
decreases a factor of $1/\mu$.  Since the slope of the galaxy number
counts is $\sim 0.17\pm 0.02$,\cite{FMDF96,Tyson88,Smail95} less
than 0.4, the galaxy number density through a cluster decreases.
Therefore the radial profile of the galaxy number counts through a
cluster is called the ``depletion curve".  The main idea to constrain the
cosmological constant by using the depletion curve, is as follows.  At
the Einstein ring radius the magnification is infinite and the
depletion curve reaches zero as its minimum value.  The Einstein ring
radius becomes larger for a higher redshift source and converges into a
finite radius for sources at the infinite redshift.  The Einstein
ring radius for  sources with infinite redshift is referred to the
``last critical line".  The observed deletion curve is the superposition
of these depletion curves of different source redshifts.  The last
critical line can be identified as the dip in the depletion curve appearing
at the largest radius.  Since the radius of the last critical line is
larger for larger cosmological constants for a fixed cluster mass
distribution model, the cosmological constant can be constrained by
identifying the last critical line for the cluster for which the mass
distribution is well constrained.  Fort, Mellier and Dantel-Fort
\cite{FMDF96} applied it to CL0024+17 and constrained the cosmological
constant to be larger than 0.6.  The merit of this method is that the
cosmological constant can be constrained with only one well-studied
cluster.  However, the cluster mass distribution model used by them
is likely to be far from reality since there is evidence of
the superposition of two large clusters in the line-of-sight,\cite{Soucail99}
and the detection of the last critical line for this cluster is also
not so clear.  A revised lens modelling and observational studies
with deeper imaging are necessary to extract a clear conclusion
on the value of the cosmological constant.

\subsubsection{GRAMORs}

Futamase, Hattori and Hamana\cite{FHH99} predicted the existence of a
highly magnified gravitationally lensed yet morphologically regular
images (they referred to as GRAMORs) in clusters of galaxies and proposed
a new method to constrain the cosmological constant by using the
GRAMORs.\cite{Hamana97b,SEIT98}  A marginally critical lens cluster may have GRAMORs if the
source galaxy is located close to the line of sight through the cluster
center.  The lowest source redshift for which the cluster can
have GRAMORs is lower for a larger cosmological constant for a fixed
cluster mass distribution model.  Clusters with intermediate X-ray
luminosities are favored to give a lower limit on the cosmological
constant, since the difference of the possible lowest source redshift
of the GRAMORs with and without the cosmological constant is largest
for those clusters.  If one identifies a very low redshift GRAMOR, it
is sufficient to give a stringent lower limit on the cosmological
constant.  The merit compared with the depletion curve method, where we
have to rely on the photometric method to measure the source redshift, is
that the redshift of the GRAMORs is spectroscopically measurable.

Although these two methods require only one cluster to constrain
the cosmological constant in principle, a precise measurement of
the cluster mass distribution is required. 
It is practically very hard.

\subsubsection{The triplet method}

Attempts to disentangle the coupling between the cosmological
parameters, the source redshift and the lens modelling have been
made by Lombardi and Bertin\cite{Lombardi98} and Gautret, Fort and
Mellier.\cite{Gautret98}  In particular, the triplet method proposed by
Gautret, Fort and Mellier \cite{Gautret98} does not require to know 
the detail of the cluster mass distribution and is summarized in
this section.  The ellipticity parameter, $\epsilon$, for a circular
source at $z_i$ is obtained by
\begin{equation}
\epsilon={\omega_i\gamma_{\infty}\over 1-\omega_i\kappa_{\infty}},\label{eq:eps}
\end{equation} 
where the subscripts $i$ refers to the redshift $z_i$,
$\gamma_{\infty}$ and $\kappa_{\infty}$ are the shear and the
convergency for the source at infinite redshift, and the ellipticity
is defined as
\begin{equation}
\epsilon={1-r\over 1+r}.\label{eq:ellip}
\end{equation}
Here $r$ is the axis ratio of the image isophotes.  Take three galaxies
with different redshifts at $z_i$, $z_j$ and $z_k$, which are located close
together in the sky through the lensing cluster and name these
``triplet".  Because of the proximity of the triplet in the sky, the
local convergency $\kappa_{\infty}$ and shear $\gamma_{\infty}$
can be assumed equal all three.  The triplet of galaxies gives 
three sets of Eq. (\ref{eq:eps}).  The quantities $\kappa_{\infty}$ and $\gamma_{\infty}$
can be deleted from these equations, and a final equation independent of
both on $\kappa_{\infty}$ and $\gamma_{\infty}$ can be derived, such as
\begin{equation}
 \left\vert
   \begin{array}{ccc}
1& \omega_i&\omega_ig_jg_k\\
1&\omega_j&\omega_jg_kg_i\\
1&\omega_k&\omega_kg_ig_j
 \end{array}
\right\vert=0.\label{eq:fin}
\end{equation} 
The final equation \label{eq:fin} is satisfied only when we 
assume correct values of the real universe for
$(\Omega_0,\Lambda_0)$.  Then the problem is reduced to determining for
the $(\Omega_0,\Lambda_0)$ values which satisfy the final equation.
Unfortunately, the method is degenerate in $(\Omega_0,\Lambda_0)$.
Therefore, a complementary cosmological constraint, such as a
high-redshift supernovae,\cite{Perlmutter98} is required to break the
degeneracy.  It is also pointed out that the triplet method is able to
measure directly the curvature of the universe, $1-\Omega_0-\Lambda_0$,
because of the mean orientation of its degeneracy.  Although in the
above discussion the source intrinsic ellipticity was omitted, errors
introduced by the intrinsic ellipticity must be taken into account.
Since the orientation and the ellipticity of the source intrinsic
shape are likely to be random, it will be canceled out if we
average over many sources.  Errors in the source redshift due to
the limitation of the photometric redshift method and systematics due
to the possible multiple lens plane, e.g., the lensing of the galaxies
at $z_j$ and $z_k$ by a galaxy at $z_i$, where $z_i$ is the lowest
within the triplet, also introduces errors in the obtained cosmological
parameters.  Therefore, applying the triplet method to many
clusters is desired.  Gautret, Fort and Mellier \cite{Gautret98} have
undertaken an intensive study on possible error estimates and have concluded
that the observation of 100 clusters can separate the $\Omega_0=0.3$
and $\Omega_0=1$ universe with a $2\sigma$ confidence level and could
be a good subject for VLT and NGST.

\section{Conclusion}

We have seen how fruitful the cluster lensing studies are.
Surprisingly, almost all areas described in this review are realizations
of Zwicky's dream.  It is no doubt that cluster lensing will
play central role as one of the standard techniques to study the
evolutionary history of the mass distribution of the universe, to probe
the deep universe, and to constrain the cosmological world models 
in the next decade with the combination of revolutionary
new observational instruments from radio to X-ray, which already have
emerged and will emerge during the next decade.

\section*{Acknowledgements}

The authors are very grateful to K. Tomita and T. Futamase for providing 
an opportunity to write this review at the right time and for their patience.
MH wishes to thank K. Hashimotodani for allowing  pre-publication of some of 
his original figures, and  thanks M. Murata for providing useful programs 
to draw caustics and time-delay surfaces.  JPK thanks  financial 
supports of CNRS-INSU and Yamada Science Foundation 
for fruitful visit to Japan. 
MH thanks financial supports of Yamada Science Foundation, and 
Grants-in-Aid by the Ministry of Education, Science and Culture of 
Japan (09740169, 60010500) in the course of the most of our studies 
summarized in this review.

\end{document}